\documentclass[trackchanges,twocolumn,twocolappendix]{aastex701}

\begin{document}

\title{An Off-Nuclear Tidal Disruption Event Discovered At Late Times: The Case Of TDE 2023mfm}

\correspondingauthor{Itai Sfaradi}
\email{itai.sfaradi@berkeley.edu}

\author[0000-0003-0466-3779]{Itai Sfaradi}
\affiliation{Department of Astronomy, University of California, Berkeley, CA 94720-3411, USA}
\affiliation{Berkeley Center for Multi-messenger Research on Astrophysical Transients and Outreach (Multi-RAPTOR), University of California, Berkeley, CA 94720-3411, USA}
\email{itai.sfaradi@berkeley.edu}

\author[0000-0002-7706-5668]{Ryan Chornock}
\affiliation{Department of Astronomy, University of California, Berkeley, CA 94720-3411, USA}
\affiliation{Berkeley Center for Multi-messenger Research on Astrophysical Transients and Outreach (Multi-RAPTOR), University of California, Berkeley, CA 94720-3411, USA}
\email{chornock@berkeley.edu}

\author[0000-0001-6747-8509]{Yuhan Yao}
\affiliation{The Kavli Institute for Astronomy and Astrophysics, Peking University, Beijing 100871, China}
\affiliation{Miller Institute for Basic Research in Science, 206B Stanley Hall, Berkeley, CA 94720, USA}
\affiliation{Department of Astronomy, University of California, Berkeley, CA 94720-3411, USA}
\affiliation{Berkeley Center for Multi-messenger Research on Astrophysical Transients and Outreach (Multi-RAPTOR), University of California, Berkeley, CA 94720-3411, USA}
\email{}

\author[0000-0002-5698-8703]{Erica Hammerstein}
\affiliation{Department of Astronomy, University of California, Berkeley, CA 94720-3411, USA}
\affiliation{Berkeley Center for Multi-messenger Research on Astrophysical Transients and Outreach (Multi-RAPTOR), University of California, Berkeley, CA 94720-3411, USA}
\email{}

\author[0000-0001-8426-5732]{Jean Somalwar}
\affiliation{Department of Astronomy, University of California, Berkeley, CA 94720-3411, USA}
\affiliation{Berkeley Center for Multi-messenger Research on Astrophysical Transients and Outreach (Multi-RAPTOR), University of California, Berkeley, CA 94720-3411, USA}
\affiliation{Kavli Institute for Particle Astrophysics and Cosmology, Stanford, CA 94305, USA}
\email{jsomalwar@berkeley.edu}

\author[0000-0003-4768-7586]{Raffaella Margutti}
\affiliation{Department of Astronomy, University of California, Berkeley, CA 94720-3411, USA}
\affiliation{Berkeley Center for Multi-messenger Research on Astrophysical Transients and Outreach (Multi-RAPTOR), University of California, Berkeley, CA 94720-3411, USA}
\affiliation{Department of Physics, University of California, 366 Physics North MC 7300, Berkeley, CA 94720, USA}
\email{rmargutti@berkeley.edu}

\author[0000-0001-8023-4912]{Huei Sears}
\affiliation{Department of Physics and Astronomy, Rutgers, the State University of New Jersey, 136 Frelinghuysen Road, Piscataway, NJ 08854-8019, USA}
\email{huei.sears@rutgers.edu}

\author[0000-0003-2434-0387]{Robert Stein}
\affiliation{Department of Astronomy, University of Maryland, College Park, MD 20742, USA}
\affiliation{Joint Space-Science Institute, University of Maryland, College Park, MD 20742, USA} 
\affiliation{Astrophysics Science Division, NASA Goddard Space Flight Center, Mail Code 661, Greenbelt, MD 20771, USA}
\email{rdstein@umd.edu}

\author[0000-0002-2249-0595]{Natalie LeBaron}
\affiliation{Department of Astronomy, University of California, Berkeley, CA 94720-3411, USA}
\affiliation{Berkeley Center for Multi-messenger Research on Astrophysical Transients and Outreach (Multi-RAPTOR), University of California, Berkeley, CA 94720-3411, USA}
\email{}

\author[00000-0003-3441-8299]{Adelle J. Goodwin}
\affiliation{International Centre for Radio Astronomy Research -- Curtin University, GPO Box U1987, Perth, WA 6845, Australia}
\email{}

\author[0000-0002-3859-8074]{Sjoert van Velzen}
\affiliation{Leiden Observatory, Leiden University, Postbus 9513, 2300 RA Leiden, The Netherlands}
\email{}

\author[0000-0002-3168-0139]{Matthew J. Graham}
\affiliation{California Institute of Technology, 1200 E. California Blvd, Pasadena, CA 91125, USA}
\email{}

\author[0000-0002-8297-2473]{Kate D.~Alexander}
\affiliation{Department of Astronomy and Steward Observatory, University of Arizona, 933 North Cherry Avenue, Tucson, AZ 85721-0065, USA}
\email{}

\author[0000-0002-8070-5400]{Nayana A. J.}
\affiliation{Department of Astronomy, University of California, Berkeley, CA 94720-3411, USA}
\affiliation{Berkeley Center for Multi-messenger Research on Astrophysical Transients and Outreach (Multi-RAPTOR), University of California, Berkeley, CA 94720-3411, USA}
\email{}

\author[0000-0002-9392-9681]{Edo Berger}
\affiliation{Center for Astrophysics \textbar{} Harvard \& Smithsonian, 60 Garden Street, Cambridge, MA 02138-1516, USA}
\email{}

\author[0000-0001-8544-584X]{Jonathan Carney}
\affiliation{Department of Physics and Astronomy, University of North Carolina at Chapel Hill, Chapel Hill, NC 27599-3255, USA}
\email{jcarney@unc.edu}

\author[0000-0001-7007-6295]{Yvette Cendes}
\affiliation{Department of Physics, University of Oregon, 1371 E 13th Ave, Eugene OR 97403, USA}
\affiliation{Institute for Fundamental Science, University of Oregon, 1371 E 13th Ave, Eugene OR 97403, USA}
\email{}

\author[0000-0003-0528-202X]{Collin~T.~Christy}
\affiliation{Department of Astronomy and Steward Observatory, University of Arizona, 933 North Cherry Avenue, Tucson, AZ 85721-0065, USA}
\email{}

\author[0000-0001-7946-1034]{Walter W. Golay}
\affiliation{Center for Astrophysics \textbar{} Harvard \& Smithsonian, 60 Garden Street, Cambridge, MA 02138-1516, USA}
\email{}

\author[0000-0002-1568-7461]{Wenbin Lu}
\affiliation{Department of Astronomy, University of California, Berkeley, CA 94720-3411, USA}
\affiliation{Theoretical Astrophysics Center, University of California, Berkeley, CA 94720, USA}
\email{}

\author[0000-0003-3124-2814]{James C. A. Miller-Jones}
\affiliation{International Centre for Radio Astronomy Research -- Curtin University, GPO Box U1987, Perth, WA 6845, Australia}
\email{}

\author[0000-0002-4557-6682]{Charlotte Ward}
\affil{Department of Astronomy and Astrophysics, 525 Davey Lab, 251 Pollock Road, The Pennsylvania State University, University Park, PA 16802, USA} 
\email{}

\author[0000-0000-0000-0000]{Eli Wiston}
\affiliation{Department of Astronomy, University of California, Berkeley, CA 94720-3411, USA}
\affiliation{Berkeley Center for Multi-messenger Research on Astrophysical Transients and Outreach (Multi-RAPTOR), University of California, Berkeley, CA 94720-3411, USA}
\email{}

\author[0000-0001-9152-6224]{Tracy X. Chen}
\affiliation{IPAC, California Institute of Technology, 1200 E. California Blvd, Pasadena, CA 91125, USA}
\email{}

\author[0000-0002-5619-4938]{Mansi M. Kasliwal}
\affil{Division of Physics, Mathematics, and Astronomy, California Institute of Technology, Pasadena, CA 91125, USA}
\email{mansi@astro.caltech.edu}

\author[0000-0002-8532-9395]{Frank J. Masci}
\affiliation{IPAC, California Institute of Technology, 1200 E. California Blvd, Pasadena, CA 91125, USA}
\email{}

\author[0000-0003-1546-6615]{Jesper Sollerman}
\affiliation{The Oskar Klein Centre, Department of Astronomy, Stockholm University, AlbaNova, SE-10691 Stockholm, Sweden}
\email{jesper@astro.su.se}

\begin{abstract}
We present the discovery and analysis of the optically selected tidal disruption event (TDE) 2023mfm, which originates from a wandering massive black hole (MBH) in a massive ($\sim 10^{11} \,$M$_{\odot}$) galaxy hosting a central low-luminosity active galactic nucleus (AGN). Our analysis of the ZTF, Lick, Keck, \emph{Swift}, Chandra, XMM-Newton, HST, and VLA observations reveals that TDE\,2023mfm is a TDE-H from a $10^{6.2 \pm 0.5}\,$M$_{\odot}$ black hole which is offset by $0.66 \pm 0.02\arcsec$ from the center of its host galaxy, corresponding to a projected distance of $1.08 \pm 0.04$\,kpc. TDE\,2023mfm displays all the traits of optically selected TDEs. The emission remains hot, $T_{\rm bb} \sim 22,000 \pm 1,000$\,K, for more than a month after peak, and the $g$-band light curve peaks at $\left( 2.24^{+0.10} _{-0.11} \right) \times 10^{43} \, \rm erg \, s^{-1}$ and stays above half-maximum luminosity for $47.8^{+3.7}_{-3.5}$ days. The late-time optical-to-UV emission observed with the HST suggests the presence of an unresolved stellar population with a stellar mass of $10^7-10^8\,$M$_{\odot}$ at the position of the TDE, possibly being a low-mass dwarf galaxy or a stripped galaxy from a previous minor merger. The radio emission from TDE\,2023mfm emerges at least a year after optical discovery, and we find it likely that it originates from a delayed outflow, however, further observations are needed to determine its true origin.
\end{abstract}


\section{Introduction}
\label{sec: intro}

Massive black holes (MBHs) are expected to reside in the centers of massive galaxies \citep{Kormendy_2013}. Observations, simulations, and theoretical frameworks suggest that a population of ``wandering'' MBHs may also exist at varying distances from galactic centers \citep{Tremmel_2018_wandering, Tremmel_2018_pair, Ricarte_2021_EM, Ricarte_2021_demographic}. These off-nuclear MBHs may arise through several channels, including gravitational-wave (GW) recoil \citep{Bekenstein_1973, Blecha_2008} or ``slingshot'' kick \citep{Ryu_2018} following MBH mergers, incomplete orbital decay during galaxy mergers \citep{Dosopoulou_2017}, or long-lived MBH binaries and triples \citep{Hoffman_2007, Tremmel_2018_pair}. 

Detecting such off-nuclear MBHs provides a unique opportunity to study the demographics and evolution of intermediate-mass black holes (IMBHs; $10^2 \, M_{\odot} \lesssim M_{\rm BH} \lesssim 10^5 \, M_{\odot}$) and MBHs ($M_{\rm BH} > 10^5 \, M_{\odot}$; \citealt{Greene_2020}). However, this is challenging because the BHs are often quiescent and lack persistent accretion signatures. One way to discover these offset MBHs is through dual and offset Active Galactic Nuclei (AGN; \citealt{Van_Wassenhove_2012, Comerford_2015, DeRosa_2019, Hogg_2021, Chen_2023}), which introduce a selection bias as these systems are already accreting. An emerging avenue to reveal the presence of otherwise dormant wandering MBHs is offered by the transient electromagnetic emission from tidal disruption events (TDEs).

TDEs occur when a star passes too close to an MBH and gets torn apart by the extreme gravitational forces \citep{Hills_1975, Rees_1988}. About half of the material remains bound and is accreted by the MBH, while the other half is unbound and can form an unbound debris stream. This results in bright emission across the electromagnetic (EM) spectrum. Originally, TDEs were discovered due to their soft X-ray emission, which is typically associated with the accretion disk formed by the stellar debris. Thanks to the growing number of all-sky optical surveys, such as All-Sky Automated Survey for SuperNovae (ASASSN; \citealt{asassn_paper}) and the Zwicky Transient Facility (ZTF; \citealt{Bellm_2019, Graham_2019, Dekany_2020}), most TDEs are now discovered through their thermal emission at optical wavelengths \citep{van_velzen_2021, Yao_2023, Hammerstein_2023}. In addition to the thermal emission observed in the optical/UV and X-ray bands, about $30\%$ of TDEs exhibit early radio emission (within the first few weeks to months after stellar disruption; \citealt{Alexander_2020}) and $\sim 40\%$ exhibit late-time radio (re)brightening (months to years after stellar disruption; \citealt{Cendes_2024}).

So far, several wandering MBHs have been discovered due to their TDE-like emission. These include the X-ray-selected TDE candidates 3XMM J215022.4-055108 (hereafter 3XMM J2150; \citealt{Lin_2018, Lin_2020}), EP240222a \citep{Jin_2025}, and the eROSITA candidate J142140.3-295325 (hereafter J1421; \citealt{Grotova_2025}). Optically selected off-nuclear TDEs include TDE\,2024tvd \citep{Yao_2025, Sfaradi_2025, Patra_2025} and TDE\,2025abcr \citep{Stein_2026, Patra_2026}. In addition, arguments in favor of an IMBH-TDE scenario have been made for EP250702a \citep{Levan_2025, Sears_2026}. While the off-nuclear location of this source is certain, the nature of EP250702a as a stellar collapse rather than a TDE is highly debated.

All of these TDE candidates have projected locations that suggest that they are bound to a massive galaxy, with stellar masses of $\sim 10^{11} \, M_{\odot}$ \citep{Guolo_2026}. On the other hand, the observational properties of the TDE\,2024tvd are unique among the other off-nuclear TDE-candidates. First, while the rest of these TDEs were found at large separations from the centers of their hosts ($\geq 5.5$\,kpc), TDE\,2024tvd resides only $\sim 0.8$\,kpc from the center of its host. In addition, TDE\,2024tvd showed a double-peaked radio light curve with an extreme temporal evolution \citep{Sfaradi_2025}. This was associated with at least one outflow (and possibly two) interacting with a complex circum-nuclear medium (CNM). In comparison, no radio emission was detected for 3XMM J2150, EP240222a, and TDE\,2025abcr. On the other hand, faint radio emission for J1241 \citep{Goodwin_2025} and a bright radio source was detected for EP250702a and was associated with a relativistic jet \citep{Goodwin_2026}; however, the TDE nature of these sources is uncertain. In this letter, we present the identification of a new off-nuclear TDE, TDE\,2023mfm.

\subsection{TDE\,2023mfm}
\label{subsec: TDE2023mfm_intro}

TDE\,2023mfm was first discovered as a transient by the ZTF \citep{Bellm_2019, Graham_2019, Dekany_2020} on 2023 June 28, and was reported to the Transient Name Server (TNS\footnote{\url{https://www.wis-tns.org/}}) on 2023 July 1 \citep{Fremling_2023_23mfm}. It was identified as a TDE candidate at a redshift of $z=0.087$ based on its blue colors ($g-r \sim -0.3$\,mag for more than $30$ days) and spectroscopically classified as a TDE based on a broad H$\alpha$ feature \citep{Chornock_2023_23mfm}. \emph{Swift} observed TDE\,2023mfm on 2023 August 10. The Neil Gehrels Swift Observatory (\emph{Swift}) Ultra-Violet/Optical Telescope (UVOT) observation revealed bright UV emission, while the X-ray observation with \emph{Swift} X-ray Telescope (XRT) resulted in a nondetection, with $L_{\rm X} < 8.8 \times 10^{41} \, \rm erg \, s^{-1}$, in the range of 0.3--10\,keV (assuming an absorbed blackbody spectral shape with $N_{\rm H}=3.31 \times 10^{20} \, \rm cm^{-2}$ and $T_{\rm bb}=80$\,eV; \citealt{Chornock_2023_23mfm_astronote}). In addition, a radio observation using the National Science Foundation's (NSF) Karl G. Jansky Very Large Array (VLA) at a central frequency of $15$\,GHz resulted in a $3\sigma$ upper limit of $F_{\rm \nu } < 0.084$\,mJy, corresponding to $\nu L_{\rm \nu} < 2.6 \times 10^{38} \, \rm erg \, s^{-1}$ \citep{Golay_2023}.

Here we present and analyze, for the first time, archival and new observations of TDE\,2023mfm. These include optical and UV photometry with the ZTF, Hubble Space Telescope (HST), and \emph{Swift} UVOT, optical spectroscopy with Keck and Lick observatories, X-ray observations with Chandra and XMM-Newton, and multi-frequency radio observations with the VLA. The off-nuclear position of TDE\,2023mfm was first suggested by \cite{Stein_2026}. However, due to the large scatter in the ZTF position and the uncertainties in the subtraction of the template, the off-nuclear position cannot be confirmed by these observations alone. An analysis of VLA observations (PI C. Christy), which establishes the off-nuclear location of TDE\,2023mfm is presented by \citet{Wenkai_2026}. Our new observations and analysis establish beyond reasonable doubt the off-nuclear position of TDE\,2023mfm and its TDE nature.

This letter is structured as follows: \S\ref{sec: observations} presents the optical and UV photometry, optical spectroscopy, and X-ray and radio observations of TDE\,2023mfm. Throughout our analysis we adopt a flat $\Lambda$CDM cosmology with $H_{0} = 70 \, \rm km \, s^{-1} \, Mpc^{-1}$, $\Omega_{m} = 0.3$, and $\Omega_{\Lambda}=0.7$. In \S\ref{subsec: tde_nature}, we discuss the TDE nature of this transient. In \S\ref{subsec: host_analysis}, we analyze the broadband (radio, optical, and UV) emission from the host galaxy. \S\ref{subsec: off_nuclear_analysis} is dedicated to the origin of the offset MBH producing the TDE, and in \S\ref{subsec: radio_outflow_analysis} we discuss the origin of the radio-emitting outflow. \S\ref{sec: conclusion} presents our conclusions.

\section{Panchromatic Observations and Analysis}
\label{sec: observations}

\subsection{Optical photometry with the ZTF}
\label{subsec: ztf}

\begin{figure}[ht]
\centering
\includegraphics[width=\linewidth]{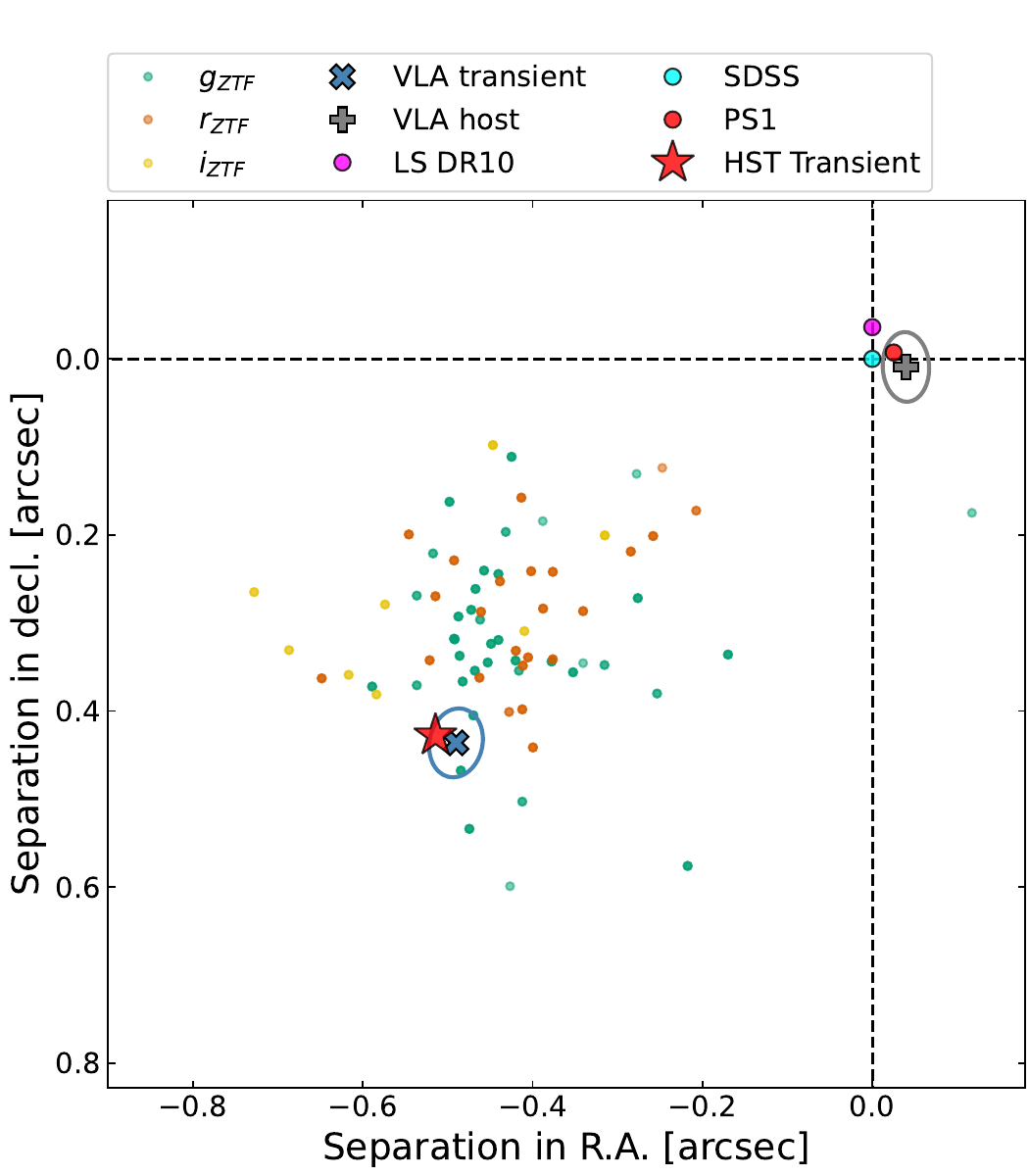}
\caption{Relative separation from the Gaia coordinates of the center of the host galaxy. The scatter plot is of the ZTF alert astrometry in $g$, $r$, and $i$-bands (see \S\ref{subsec: optical_photometry}). The sources detected with the VLA of the host nucleus (grey) and the TDE (blue) are plotted with a region that represents $2\sigma$ uncertainty of the beam position (see \S\ref{subsec: vla}). The TDE position obtained from the HST/WFC3 F275W image is marked with a red star. Also plotted for reference is the position of the host in Legacy Survey Data Release 10 (LS DR10), Sloan Digital Sky Survey (SDSS), and Pan-STARRS1 (PS1).
\label{fig: positional_analysis_plot}}
\end{figure}

We obtain ZTF forced photometry in the $g$, $r$, and $i$ bands \citep{Masci_2019, Masci_2023} based on the median position of the ZTF alerts reported in Fritz \citep{van_der_Walt2019, Coughlin_2023}: R.A.: 21$^{\rm h}$37$^{\rm m}$27$^{\rm s}$.96, Decl.: $-04^{\rm \circ}$20$^{\rm '}$43$^{\rm "}$.02 (J2000; see the scatter of the alert astrometry in Fig.~\ref{fig: positional_analysis_plot}). We then apply the quality cuts presented in \S6.1 of \cite{Masci_2023} and a baseline flux correction. The baseline correction is applied by manually choosing a time window prior to the optical flare of TDE\,2023mfm and visually inspecting the light curve to remove uncertainties regarding excess emission. For each band and field we calculate the average flux density in the given time window and remove it from the entire light curve\footnote{We provide the code used for this analysis at \url{https://github.com/itaisfaradi/ZTF_FP_Reduction}}. We apply an extinction correction based on the Galactic color excess $E \left(B-V\right) = 0.0342 \pm 0.0003$\,mag estimated by \cite{Schlafly_2011} and the extinction law presented in \cite{Cardelli_1989} with $R_{V} = 3.1$. The baseline and extinction corrected light curves are plotted in Fig.~\ref{fig: ZTF_UVOT} and reported in Table~\ref{tab: optical_photometry} in Appendix~\ref{sec: table_appendix}.

\begin{figure}
\centering
\includegraphics[width=\linewidth]{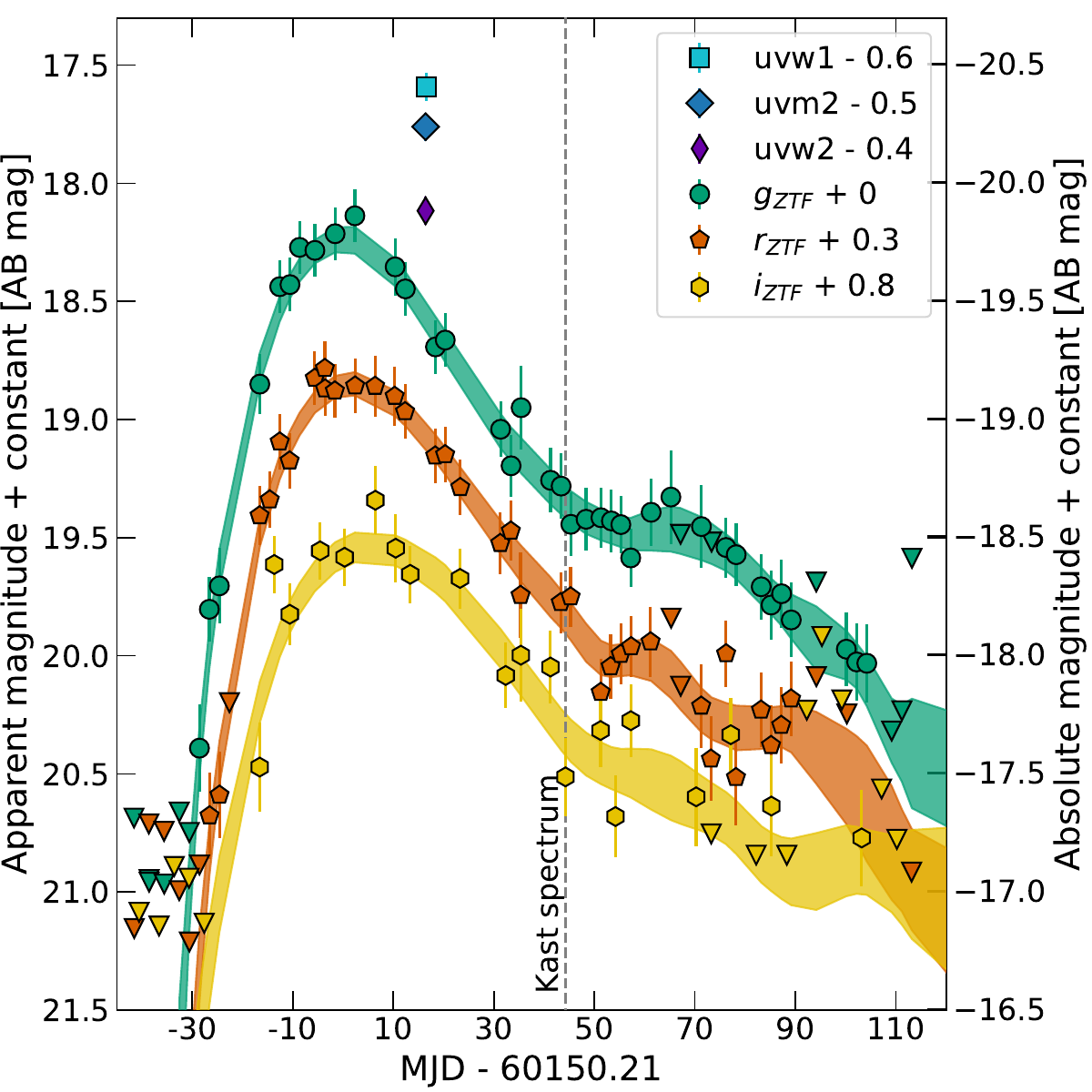}
\caption{Optical (bands $g$, $r$, and $i$ of the ZTF) and UV (\emph{Swift} UVOT bands \emph{uvw1}, \emph{uvw2}, and \emph{uvm1}) light curves of TDE\,2023mfm. $5\sigma$ upper limits are plotted with triangles. The optical light curves are corrected for baseline emission and Galactic extinction as described in \S\ref{subsec: ztf}. The UV emission is after subtraction of the host galaxy emission and Galactic extinction correction (see \S\ref{subsec: swift_uv}). Also plotted as colored-regions are $1\sigma$ regions around the best-fitting Gaussian-processes to the optical light curves (see \S\ref{subsec: optical_photometry}). We mark the time the Kast spectrum was obtained (see \S\ref{subsec: optical_spectroscopy}) with a vertical dashed line. For clarity, we offset the light curves in this plot by a different constant (which is specified in the legend) for each band. 
\label{fig: ZTF_UVOT}}
\end{figure}

\subsection{UV photometry with Swift/UVOT}
\label{subsec: swift_uv}

TDE\,2023mfm was followed up with the Neil Gehrels Swift Observatory \citep{Gehrels_2004} in the UV with Swift/UVOT \citep{Roming_2005}. We used the \texttt{uvotsource} package to analyze the Swift/UVOT photometry, using an aperture of $5"$. We used the stellar population synthesis described in \S\ref{subsec: host_analysis} to estimate the host galaxy flux in the UVOT bands and subtract it. We report the Swift/UVOT host-subtracted and extinction corrected photometry measurements in Table~\ref{tab: swift_uvot_photometry} in Appendix~\ref{sec: table_appendix}.

\subsection{Analysis of the Early-Time Optical/UV Photometry}
\label{subsec: optical_photometry}

We interpolated the light curve in the ZTF bands using Gaussian Processes (GPs), a non-parametric regression technique that enables the reconstruction of the light-curve evolution between observations. We used \texttt{Scikit-learn}'s implementation \citep{scikit-learn} with a \texttt{Matern} kernel and a length scale of $0.05$ for both the $g$ and $r$ bands, and $0.5$ for the $i$ band. We present the results in Fig.~\ref{fig: ZTF_UVOT}, from which we infer the time of the peak of the $g$-band light curve to be MJD $60150.2^{+2.3}_{-2.1}$ ($\delta t$ is defined hereafter as the time since the $g$-band peak). These reconstructed light curves are used to infer the observational parameters presented in \S~\ref{subsec: tde_nature} and to compare them with those of other known TDEs. We also interpolate the resulting flux density observations in the $g$ and $r$ bands to the time of the \emph{Swift}/UVOT observations (MJD $= 60166$; $\delta t = 16$\,d) and construct a broadband spectral energy distribution (SED).

We fit the optical-to-UV SED at this epoch with a single-temperature blackbody model. We implemented Markov chain Monte Carlo (MCMC) using \texttt{emcee} \citep{Foreman_Mackey_2013}, with the blackbody temperature, $T_{\rm bb}$, and radius, $R_{\rm bb}$, as free parameters. We use $100$ walkers for $2000$ steps per walker and discard the first $200$ steps for burn-in. We confirm convergence and mixing across both parameters by obtaining a small potential scale reduction factor ($\hat{R}\simeq1.01$ for both parameters), a high Effective Sample Size (${\rm ESS} \gtrsim 6200$ for both parameters), and short integrated autocorrelation times ($\tau \simeq 28$ for both parameters). The best-fitting parameters are $\log_{10} \left(T_{\rm bb} / {\rm K} \right) = 4.26 \pm 0.01$ and $\log_{10} \left( R_{\rm bb} / {\rm cm} \right) = 14.93 \pm 0.02$.

\subsection{Optical Spectroscopy}
\label{subsec: optical_spectroscopy}

\begin{figure}
\centering
\includegraphics[width=\linewidth]{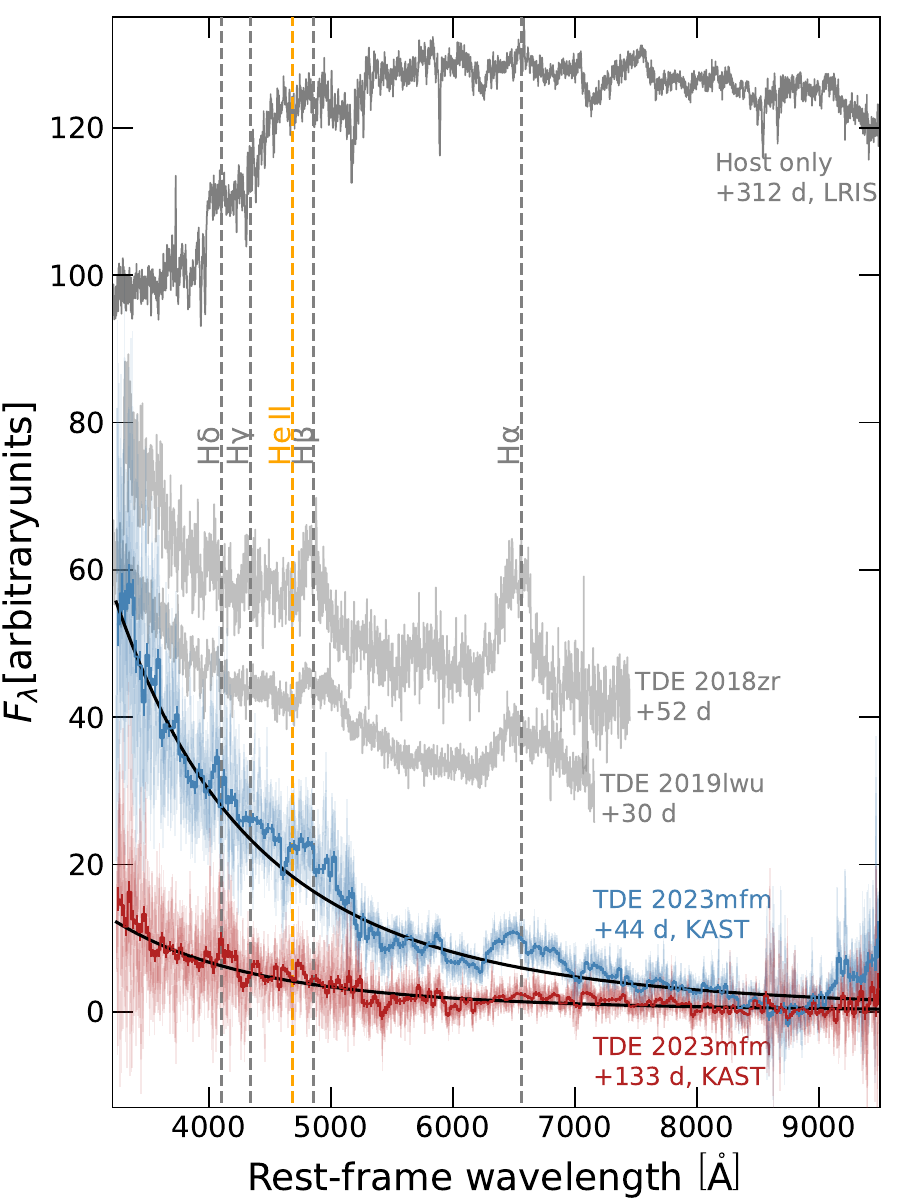}
\caption{Optical spectra of TDE\,2023mfm obtained $44$ days (blue) and $133$ days (red) after the $g$-band peak. In black we plot the optical spectra of the host obtained $312$ days after peak, and after the transient emission has faded. The spectra at $\delta t = 44$\,d and $133$\,d are host-subtracted, and modeled with a blackbody continuum (plotted with black solid lines) as described in \S~\ref{subsec: optical_spectroscopy}. We also plot (in gray) archival spectra of two TDE-H (AT\,2018zr and AT\,2019lwu) presented in \cite{Hammerstein_2023} for reference. The broad H$\alpha$ and H$\beta$ emission features seen in the spectrum on $\delta t = 44$\,d are used to spectroscopically classify TDE\,2023mfm as a TDE-H. Finally, we emphasize here that the spectrum obtained at $\delta t = 133$\,d still presents clear evidence for a blue continuum, strengthening the arguments against a SN origin for TDE\,2023mfm.
\label{fig: optical_spectroscopy}}
\end{figure}

We obtained two optical spectra of TDE\,2023mfm with the Kast double spectrograph on the Shane $3$\,m telescope at Lick Observatory \citep{Miller_1994} on 2023 September 7 and 2023 December 5, corresponding to $\delta t \simeq 44$\,d and $133$\,d, respectively (PI R. Chornock). In addition, we obtained one spectrum with the Keck Low-Resolution Imaging Spectrometer (LRIS; \citealt{Oke_1995}) on 2024 June 1 ($\delta t \simeq 312$\,d; PI R. Chornock). This spectrum did not show any transient emission, and we used it to model the emission from the host. These spectra are plotted in Fig.~\ref{fig: optical_spectroscopy}. In addition, four optical spectra were obtained with the Spectral Energy
Distribution Machine (SEDM; N. \citealt{Blagorodnova_2018, Rigault_2019, Kim_2022}) on
the robotic Palomar 60 inch telescope (P60; \citealt{Cenko_2006}). These spectra are available as an online supplementary files.

We next fitted the spectra with the transient emission to remove the underlying emission from the host. We matched the spectral resolution of the LRIS and Kast spectra by convolving the higher-resolution spectrum with a Gaussian kernel. We modeled the transient spectra at $\delta t = 44$\,d and $133$\,d with a scaled host spectrum plus a scaled blackbody with a temperature $T_{\rm bb}$. While fitting, we also corrected for Galactic extinction in the model and masked Balmer lines. The blackbody temperature was fitted as a free parameter for the spectrum at $\delta t = 44$\,d. For the spectrum at $\delta t = 133$\,d we fixed the blackbody temperature to the value inferred from the fit to the spectrum at $\delta t = 44$\,d.

We present the host-subtracted spectra with the best-fitting blackbody in Fig.~\ref{fig: optical_spectroscopy}. We first note that both spectra at $\delta t = 44$\,d and $133$\,d exhibit blue continuum emission on top of the host galaxy emission. The best-fitting blackbody temperature from the fit to the spectrum on $\delta t=44$\,d is $T_{\rm bb} = \left( 2.2 \pm 0.1 \right) \times 10^{4} \, \rm K$. In addition, the spectrum at $\delta t = 44$\,d exhibits a broad H$\alpha$ line (with line velocity width of $\sim 18,000 \, \rm km \, s^{-1}$) and, possibly, a broad H$\beta$ line, similar to other known TDE-H (see examples plotted in Fig.~\ref{fig: optical_spectroscopy}). We do not detect these lines in the spectrum from $\delta t = 133$\,d. A more detailed discussion of this spectral classification in the context of the TDE nature of TDE\,2023mfm is presented in \S\ref{subsec: tde_nature}.

\subsection{Late-Time Optical/UV Photometry with HST}
\label{subsec: hst}

\begin{figure*}[ht]
\centering
\includegraphics[width=\linewidth]{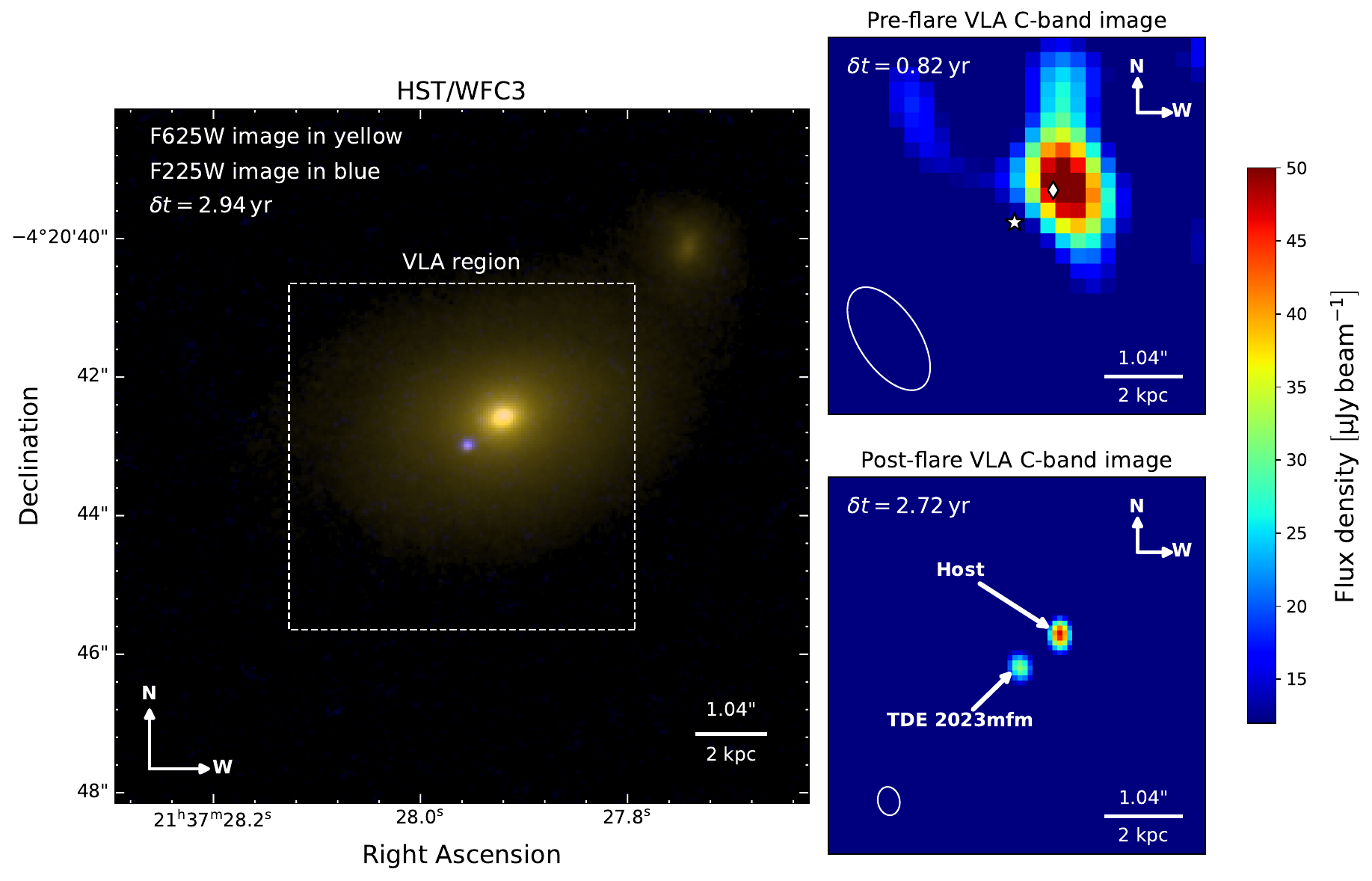}
\caption{Images of the field of TDE\,2023mfm in optical, UV, and radio wavelengths. \textbf{Left panel}: HST/WFC3 images with F225W and F625W filter (blue and yellow, respectively). \textbf{Top right}: VLA image in C-band (with a central frequency of $6$ GHz) before the turn on of the radio emission from the TDE, and while the VLA was in B-configuration. \textbf{Bottom right}: VLA image in C-band after the turn on of the radio emission from the TDE, and while the VLA was in A-configuration. The positions of the host nucleus and the TDE (obtained from the HST F625W image) are marked with a diamond marker and a star marker, respectively. The scale of the flux level of both radio images is the same. The pre-flare radio image was taken in the VLA B-configuration and the post-flare image was taken in A-configuration, resulting in different angular resolution as seen in the size of the clean beams plotted as white ellipses at the bottom-left of these images. However, it is still clear that the emission at the position of the host nucleus is roughly constant (probably due to a low-luminosity AGN as discussed in \S~\ref{subsec: host_analysis}) and that the source at the position of TDE\,2023mfm is from an off-nuclear transient source.
\label{fig: optical_radio_images}}
\end{figure*}

We obtained HST imaging on 2026 July 2 ($\delta t = 1073$\,d) under our dedicated program for off-nuclear TDEs using the Wide Field Camera 3 (WFC3) with bands F225W, F275W, F336W, and F625W (the data described here may be obtained from the MAST archive at \dataset[doi:10.17909/fn8k-3y84]{https://dx.doi.org/10.17909/fn8k-3y84}). All four images show a point source at the position of TDE\,2023mfm, and an extended source at the position of the host galaxy. We present a color image composed of F225W and F625W images in Fig.~\ref{fig: optical_radio_images}.

We use \texttt{GALFIT}, while fitting for the centers of both the transient and the host, to determine their positions and measure the emission from them. Since there are only a few faint stars in the field we use the publicly available\footnote{\url{https://www.stsci.edu/~jayander/HST1PASS/LIB/PSFs/STDPSFs/}} PSF for each band (since there is no available PSF for F625W we used the F621W instead). The F225W, F275W and F336W images were modeled with a single point source for TDE\,2023mfm, and with a S\'{e}rsic profile for the host. The morphology of the host in the F625W image is more complicated, we therefore modeled it as a S\'{e}rsic profile and an exponential disk model. There are additional nearby galaxies in the F625W image, and we modeled the most nearby one (the north-west source seen in Fig.~\ref{fig: optical_radio_images}) with a S\'{e}rsic profile. These images and their fits are presented in Fig.~\ref{fig: HST_GALFIT_images} in Appendix \ref{sec: plots_appendix}. We note here that due to a systematic $\sim 0.1\arcsec$ difference between the center of the host observed in the HST and all other available images of the host galaxy (including Legacy Survey, SDSS, PS1, Gaia, and the VLA), we assume that the position of the center of the host is at the position of the source seen in Gaia. Therefore, we correct the absolute astrometry of the HST images by shifting their coordinates such that the center of the host will match its Gaia position.

Based on our \texttt{GALFIT} fit to the F275W image we find that the position of the source at the center of the galaxy is R.A.: 21$^{\rm h}$37$^{\rm m}$27$^{\rm s}$.9274 $\pm 0.0013 \, \rm s$, Dec.: $-04^{\rm \circ}$20$^{\rm '}$42$^{\rm "}$.709 $\pm 0.015^{\rm "}$ (J2000), and at the position of the TDE it is R.A.: 21$^{\rm h}$37$^{\rm m}$27$^{\rm s}$.9613 $\pm 0.0014 \, \rm s$, Dec.: $-04^{\rm \circ}$20$^{\rm '}$43$^{\rm "}$.129 $\pm 0.016^{\rm "}$ (J2000). This translates to a separation of $0.66 \pm 0.02\arcsec$, and at a redshift of the host galaxy, $z=0.087$, the projected distance between the center of the host and the off-nuclear TDE is $1.08 \pm 0.04$\,kpc. We report the photometry measurements of TDE\,2023mfm in Table~\ref{tab: hst_photometry} in Appendix \ref{sec: table_appendix} and plot them in Fig.~\ref{fig: late_time_phot}.

\subsection{X-ray observations with Chandra}
\label{subsec: chandra}

TDE\,2023mfm was observed by the Chandra X-ray Observatory with the Advanced CCD Imaging Spectrometer (ACIS; \citealt{Garmire2003}) under a GO program on 2026 May 23 for a total of 52.0\,ks (obsID 31222). We processed the data using the Chandra Interactive Analysis of Observations (\texttt{CIAO}; \citealt{Fruscione2006}) software package (v4.18). 

\subsubsection{Source profile and location} \label{subsubsec:cxo_position}

A faint X-ray source is detected around the location of the host galaxy. To assess whether the source is extended, we simulated the PSF using the Chandra Ray Tracer (ChaRT; \citealt{Carter2003}), projected the PSF onto the detector-plane with MARX \citep{Davis2012}, and created an image of the PSF. Using \texttt{srcextent}, we estimated that the observed source size is 0.95$^{\prime\prime}$ (90\% confidence interval: 0.46--1.44$^{\prime\prime}$), the PSF observed size is 0.46$^{\prime\prime}$ (90\% confidence interval: 0.46--0.47$^{\prime\prime}$). The source is not extended at 90\% confidence.

To determine the astrometric accuracy of the Chandra image, we first ran \texttt{fluximage} to create a 1--7\,keV image and then applied \texttt{wavdetect} to identify point sources on the ACIS-S S2 and S3 chips. 
We cross-matched the resulting source catalog with Gaia DR3\footnote{The Gaia source associated with the host-galaxy nucleus was excluded from this cross-matching procedure.} using a matching radius of $2\arcsec$, yielding six Chandra/Gaia matches. We then used \texttt{wcs\_match} to determine whether a global astrometric correction could improve the image registration. The fitted translational solution did not reduce the overall positional residuals. We therefore applied no astrometric correction and adopted a nominal (1$\sigma$) systematic astrometric uncertainty of $\sigma_{\rm sys} = 0\farcs69$ for ACIS-S\footnote{\url{https://cxc.cfa.harvard.edu/mta/ASPECT/celmon/}}. The X-ray source in the vicinity of TDE\,2023mfm found by \texttt{wavdetect} is at ${\rm R.A.} = 21^{\rm h} 37^{\rm m} 27.93^{\rm s}$, ${\rm decl.} = -04^{\circ} 20^{\prime} 42.59^{\prime\prime}$ (J2000), with a statistical uncertainty of 0\farcs16. Combining systematic and statistical errors, the $1\sigma$ uncertainty is 0\farcs71. 

The Chandra source is 0\farcs14 from the HST host nucleus and  0\farcs68 from the HST transient location (see \S\ref{subsec: hst}). Using the method outlined in \citet{Yao_2025}, we calculate the posterior probability that the X-ray source is associated with each position. We adopt equal prior probabilities for association with the host nucleus and the transient and evaluate a two-dimensional Gaussian positional likelihood, including the 0\farcs71 Chandra astrometric uncertainty. We obtain association probabilities of 61\% and 39\% for the host nucleus and the transient, respectively. Thus, the Chandra position slightly favors an association with the host nucleus.

\subsubsection{Source spectrum} \label{subsubsec:cxo_spec}

We extracted the source spectrum using a source region of $r_{\rm src}=1.5^{\prime\prime}$ centered on the X-ray position determined by \texttt{wavdetect}. The background spectrum was extracted using nearby source-free regions. We grouped the Chandra spectrum to at least one count per bin, and modeled the 0.5--7\,keV data using $W$-statistics. A total of (0.5--7\,keV) 11 counts were detected within the source region. 
Using a model of \texttt{tbabs*cflux*powerlaw} and fixing the Galactic column density at $N_{\rm H}=3.9\times 10^{20}\,{\rm cm^{-2}}$, we obtained a best-fit model with \texttt{Wstat/dof=4.65/9} and $\Gamma = 2.14\pm0.84$. The unabsorbed 0.3--10\,keV flux is ${\rm log}[f_{\rm X}/({\rm erg\,s^{-1}\,cm^{-2}})] = -14.13_{-0.08}^{+0.24}$, corresponding to an X-ray luminosity of $1.4\times 10^{41}\,{\rm erg\,s^{-1}}$. 

The X-ray spectral index is more similar to those observed in active galactic nuclei (AGNs) than in TDEs \citep{Guolo2024}. Combined with the astrometric information presented in \S\ref{subsec: hst}, this favors an association of the X-ray source with the host-galaxy nucleus. The measured nuclear X-ray luminosity is much brighter than what is expected from X-ray binaries \citep{Zou2025}. The Eddington ratio ($\lambda_{\rm Edd}\equiv k_{\rm bol}L_{\rm X} / L_{\rm Edd}$) is then ${\rm log}\lambda_{\rm Edd} \approx -4$, where we take $k_{\rm bol}=25$ following \citet{Zou2025}. This is consistent with a nuclear low-luminosity AGN.

\subsubsection{TDE X-ray upper limit}
\label{subsubsed: x_ray_limit_tde}

Given the non-detection of TDE\,2023mfm by Chandra, we estimated the background-limited image sensitivity and corresponding X-ray upper limit. Using a source-free region, we found that the expected 0.3--8\,keV background contribution within a circular aperture of radius \(2\arcsec\) is \(\lambda_{\rm b}=0.782\) counts. For a Poisson distribution with mean \(\lambda_{\rm b}\), the minimum one-sided \(3\sigma\) detection threshold is five counts; the probability of obtaining at least five counts from the background alone is \(1.28\times10^{-3}\). Requiring a 90\% probability of exceeding this threshold gives a limiting expected source contribution of 7.21 counts within the aperture. The \(2\arcsec\) aperture encloses 93.66\% of the point-spread function, yielding a PSF-corrected limit of 7.70 counts. For an effective exposure time of 49.42 ks, this corresponds to a limiting count rate of \(1.56\times10^{-4}\ {\rm count\,s^{-1}}\). Adopting an absorbed power-law spectrum with $N_{\rm H} = 3.9\times 10^{20}\,{\rm cm^{-2}}$ and $\Gamma = 4$, the 0.3--10\,keV flux limit is $f_{\rm X} < 1.72\times 10^{-14}\,{\rm erg\,s^{-1}\,cm^{-2}}$, which corresponds to $L_{\rm X} < 3.24\times 10^{41}\,{\rm erg\,s^{-1}}$. We discuss the implications of this non-detection in \S\ref{subsec: tde_nature}.

\subsection{X-ray observations with XMM-Newton}
\label{subsec: xmm}

We observed the field of TDE\,2023mfm with the XMM-Newton telescope under AO-24 program 96333 (PI R. Chornock) on 2026 June 1 for a total of 31.4\,ks. We reduced the EPIC pn data using the XMM-Newton Science Analysis System\footnote{\href{https://www.cosmos.esa.int/web/xmm-newton/sas}{https://www.cosmos.esa.int/web/xmm-newton/sas}} (SAS) following standard procedures. We extracted the source using a circular region with a radius of $30^{\prime\prime}$ centered on the ZTF alert median position, and the background from a source-free region on the same CCD. The MOS data are shallower than the pn data, so we omit reporting them in this paper. The extracted spectrum was binned with \texttt{specgroup} to have a mininum of three counts per bin and an oversample of three. 

Modeling the spectrum using the same model and fit statistics as in \S\ref{subsubsec:cxo_spec}, the best-fit result gives \texttt{Wstat/dof = 31.12/32}, $\Gamma = 1.43_{-0.66}^{+0.76}$, and an unabsorbed 0.3--10\,keV flux of ${\rm log}[f_{\rm X}/({\rm erg\,s^{-1}\,cm^{-2}})] = -14.14_{-0.31}^{+0.27}$, consistent with the Chandra result, and therefore, consistent with being emission from the galactic nucleus and not the TDE.

\subsection{Radio observations with the VLA}
\label{subsec: vla}

Early radio observations of AT \,2023mfm with the VLA resulted with an upper limit of $F_{\rm \nu} \leq 0.084$\,mJy at $15$\,GHz \citep{Golay_2023}. We have identified two radio sources in VLA observation (VLA 25A-167; PI C. Christy) at $\delta t = 991$\,d as part of our effort to confirm the nature of optically identified off-nuclear candidates of \cite{Stein_2026} \citep{Wenkai_2026}.

We observed the field of TDE\,2023mfm with S ($3$\,GHz), C ($6$\,GHz) and, X ($10$\,GHz) bands on three epochs, starting 2026 April 27 ($\delta t = 1007$), while the VLA in A-configuration, under our dedicated programs to observe off-nuclear TDEs (VLA 25B-109; PI I. Sfaradi and XMM-Newton/VLA \#96333; PI R. Chornock). We used 3C48 as the absolute flux and bandpass calibrator, J2134-0153 as an interleaved phase calibrator, and we used the Common Astronomy Software Applications (CASA; \citealt{CASA}) packages and the VLA calibration pipeline (v6.5.4.9) to flag and calibrate the data. To image the field of TDE\,2023mfm we used the CASA task \texttt{TCLEAN}, and \texttt{IMSTAT} was used to calculate the image rms. When available, the point sources at the position of TDE\,2023mfm and of the nucleus of the host were fitted using CASA task \texttt{IMFIT}. We estimate the uncertainty on the flux density as the square root of a quadratic sum of the uncertainty obtained by \texttt{IMFIT} and $10\%$ calibration error. Finally, the position of the point sources is also determined by the best-fitting parameters obtained by \texttt{IMFIT}, and the positional uncertainty is estimated to be $10\%$ of the synthesized beam FWHM\footnote{\url{https://science.nrao.edu/facilities/vla/docs/manuals/oss/performance/positional-accuracy}}.

In addition, we re-analyzed the observation reported by \cite{Golay_2023} to achieve better sensitivity, and an observation obtained in C-band on 2024 June 20, while the VLA was in B-configuration (VLA 20B-337; PI K. D. Alexander). The calibration and imaging were done in the same way as above. All available radio measurements of the transient source and of the source at the position of the host nucleus are reported in Tables~\ref{tab: tde_radio_observation} and \ref{tab: host_radio_observation} in Appendix~\ref{sec: table_appendix}. Finally, we provide images of the field of TDE\,2023mfm before and after the late-time radio brightening in Fig.~\ref{fig: optical_radio_images}. These pre-flare image was obtained in B-configuration and the post-flare image in A-configuration, resulting in different angular resolutions.

\section{Discussion}
\label{sec: discussion}

We start by establishing the TDE nature of this transient and compare its observational properties to a large sample of TDEs in \S\ref{subsec: tde_nature}. Then, we discuss the properties of the host galaxy of TDE\,2023mfm in \S\ref{subsec: host_analysis}, and the astrophysical origin of the two MBHs at the center of the host galaxy in \S\ref{subsec: off_nuclear_analysis}. Finally, in \S\ref{subsec: radio_outflow_analysis} we analyze the late-time brightening of the radio emission from TDE\,2023mfm and explore the possible scenarios for the emergence of the radio emitting outflow at late-times.

\subsection{The TDE Nature of TDE\,2023mfm}
\label{subsec: tde_nature}

\begin{figure*}[ht]
\centering
\includegraphics[width=\linewidth]{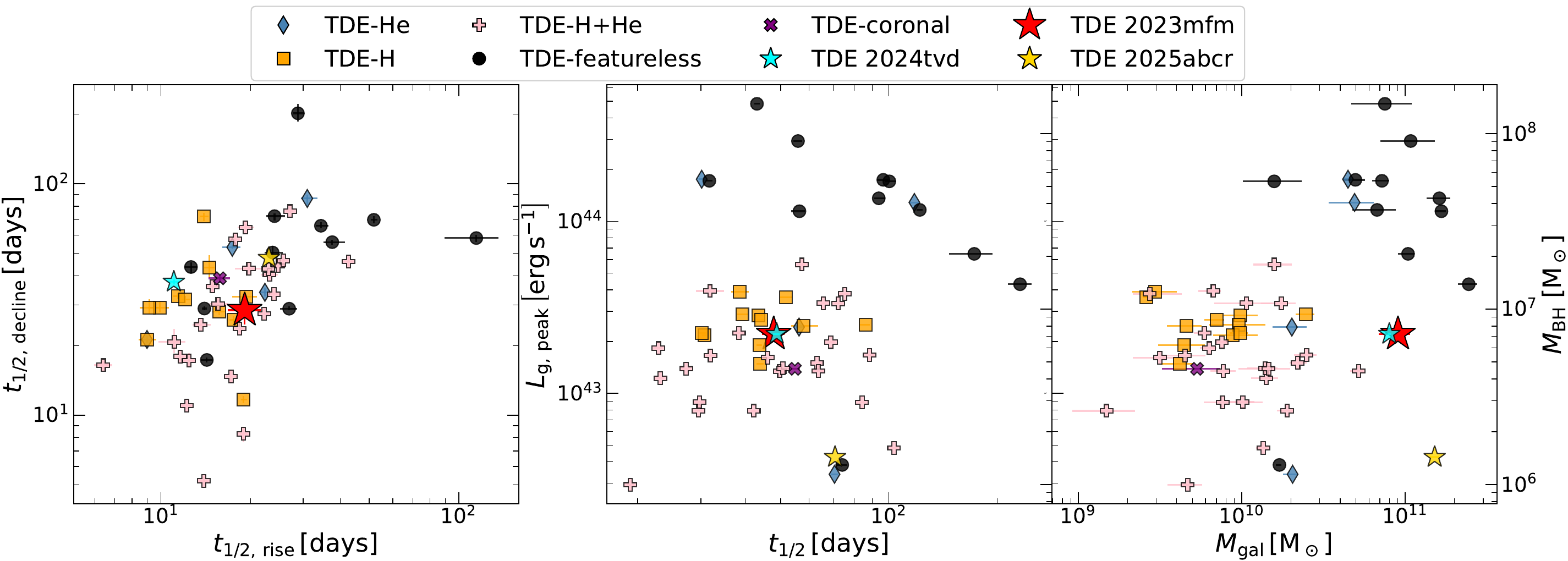}
\caption{The observational and physical properties of TDE\,2023mfm (marked with a red star) and its host, compared to previously classified TDEs (adopted from \citealt{Yao_2025}). The rise time from half-maximum to maximum $t_{\rm 1/2, \, rise}$, the decline from maximum to half-maximum, $t_{\rm 1/2, \, decline}$, the time above half-maximum, $t_{1/2}$, and the $g$-band peak luminosity, $L_{\rm g, \, peak}$, are all calculated from the reconstruction of the $g$-band light curve using GP (see \S~\ref{subsec: optical_photometry}). The mass of the host galaxy is estimated by the fit of the galaxy profile presented in \S~\ref{subsec: host_analysis}. The mass of the black hole is estimated by using the relation between $L_{\rm g, peak}$ and $M_{\rm BH}$ presented in \cite{Mummery_2024} (Eq. 74). For comparison, the other optically-selected off-nuclear TDEs, TDE\,2024tvd and TDE\,2025abcr, are plotted with cyan and golden stars, respectively. As seen from this plot, the $g$-band peak luminosity, $L_{\rm g, \, peak}$, and the time above half-maximum, $t_{1/2}$, and the mass of the host galaxy are all similar to the values inferred from TDE\,2024tvd.
\label{fig: properties_comparison}}
\end{figure*}

The optical and UV emission of TDE\,2023mfm presented in \S\ref{subsec: swift_uv}, \ref{subsec: optical_photometry}, and \ref{subsec: optical_spectroscopy}, are consistent with many of the known properties of TDEs. The optical light curve rises to peak in $\sim 30$ days and then slowly declines for $\sim 100$ days (see Fig.~\ref{fig: ZTF_UVOT}). The optical colors remain blue ($g-r \leq -0.2$\,mag) throughout its entire evolution. In addition, the blackbody temperature we infer from both the blackbody fits to the optical/UV spectral energy distribution (SED), and to the host-subtracted spectra, remains hot ($T_{\rm bb} \sim 2 \times 10^{4} \, \rm K$) more than a month after peak. Since the optical and UV emission from supernovae tend to cool significantly after the peak of the light curve, the observational properties from TDE\,2023mfm allow us to rule out a supernova origin for this transient.

Unlike in the optical and the UV bands, TDE\,2023mfm is not detected in the X-ray at both early ($\delta t \simeq 16$\,d) and late ($\delta t \simeq 3$\,yr) times. Early X-ray emission from TDEs can span a wide range of luminosities and remains below $10^{42} \, \rm erg \, s^{-1}$ \citep{Guolo2024}. Therefore, the lack of detected X-ray emission at early times, with $L_{X} < 9 \times 10^{41} \, \rm erg \, s^{-1}$ at $\delta t = 16$\,d, is consistent with many TDEs. On the other hand, a significant number of optically discovered TDEs exhibit late-time brightening of their X-ray emission, and \cite{Guolo2024} showed that about $40\%$ of the optically discovered TDEs become X-ray loud ($L_{X} \geq 10^{42} \, \rm erg \, s^{-1}$) eventually. TDE\,2023mfm remains faint ($L_X < 3.2 \times 10^{41} \, \rm erg \, s^{-1}$) at late times. However, this is not unprecedented as the luminosity of some optically selected TDEs can be as faint as $L_{X} \sim 10^{41} \, \rm erg \, s^{-1}$ years after stellar disruption \citep{Jonker_2020}.

In Fig.~\ref{fig: properties_comparison} we compare the optical properties of TDE\,2023mfm with other known TDEs. The rise time from half-maximum to maximum is $t_{\rm 1/2, \, rise} = 19.1^{+2.5} _{-2.3}$\,d and the decline from maximum to half-maximum is $t_{\rm 1/2, \, decline} = 28.7^{+3.7} _{-4.0}$\,d, similar to other TDE-H and TDE-H+He. Interestingly, while the $t_{\rm 1/2, \, decline}$ and $t_{\rm 1/2, \, rise}$ of TDE\,2023mfm are different from the other two optically-selected off-nuclear TDEs (TDE\,2024tvd and TDE\,2025abcr), the $g$-band peak luminosity, $L_{\rm g, \, peak} = \left( 2.24^{+0.10} _{-0.11}\right) \times 10^{43} \, \rm erg \, s^{-1}$, and the time above half-maximum, $t_{1/2} = 47.8^{+3.7} _{-3.5}$, are remarkably similar to the off-nuclear TDE\,2024tvd. In addition, the mass of the host we infer in \S~\ref{subsec: host_analysis} is very similar to the mass of the host of TDE\,2024tvd. On the other hand, the other optically selected off-nuclear TDE\,2025abcr exhibits a lower peak $g$-band luminosity, longer time above half-maximum, and resides in a slightly more massive galaxy. Based on all of the above, the clear broad hydrogen line in the host-subtracted optical spectrum (see \S~\ref{subsec: optical_spectroscopy}), and the off-nuclear position observed both in the ZTF alerts and in multiple high-resolution optical, UV, and radio images, we conclude that TDE\,2023mfm is an off-nuclear TDE-H. 

There are several methods to estimate the mass of the black hole using the observed thermal emission from the TDE. We first use the scaling relations with the late-time UV plateau luminosity \citep{Mummery_2024}. Using the luminosity observed with the HST filter F225W and applying Eq. 56 in \cite{Mummery_2024}, we estimate $M_{\rm BH} = 10^{6.18 \pm 0.52} \, M_{\rm \odot}$ (where we account for the intrinsic scatter of $0.5$\,dex). \cite{Mummery_2024} also showed that the mass of the black hole scales with the peak of the $g$ band luminosity. We use Eq. 74 from \cite{Mummery_2024} and the peak $g$-band luminosity we infer using the GPs to estimate $M_{BH} = 10^{6.86 \pm 0.55} \, M_{\rm \odot}$ (where we accounted for the intrinsic scatter of $0.53$\,dex). In addition to these two approaches, previous works estimated the mass of MBHs giving rise to TDEs using the Modular Open-Source Fitter for Transients (\texttt{MOSFIT}; \citealt{Guillochon_2018, Mockler_2019}). \texttt{MOSFIT} models the multi-band light curves by combining the fallback rate of tidally disrupted stellar debris onto the black hole with a viscous accretion delay and reprocessing of the accretion luminosity into thermal blackbody emission. The model parameters are constrained using MCMC sampling. We fit the optical and UV emission (up to $\delta t = 100$\,d) with \texttt{MOSFIT} and estimate a black hole mass of $M_{\rm BH} = 10^{6.5 \pm 0.1} \, M_{\odot}$. Finally, we note that \cite{Guolo_2025} recovered the mass of MBHs by fitting a fully relativistic disk model to X-ray and UV SEDs of selected TDEs. However, due to the lack of late-time soft X-ray emission from TDE\,2023mfm we cannot apply this method here.

Overall, we conclude that the mass of the black hole that caused the TDE is $\sim 10^{6}-10^{7} \, M_{\odot}$. This is similar to many of the known TDEs (see Fig.~\ref{fig: properties_comparison}).

\subsection{The host galaxy of TDE\,2023mfm}
\label{subsec: host_analysis}

\subsubsection{Galaxy profile}
\label{subsec: galaxy_profile}

Recent major (mass ratio of $1:1$) or minor ($\lesssim 1:3$) mergers will result in tidal features and disturbed morphologies in galaxies on large $\sim$kpc scales. \citet{Lotz_2008} showed that strong quantitative asymmetry (defined as $A \geq 0.35$, where A is the asymmetry measure\footnote{The asymmetry measure quantifies the degree to which the light distribution of the galaxy is symmetric (see Eq. 3 in \citealt{Lotz_2008}).}) decreases rapidly ($\sim 0.1$ to $0.5$\,Gyr) after final coalescence of equal mass gas-rich disk mergers. On the other hand, faint tidal features can remain visible for up to $\sim 1$\,Gyr after coalescence \citep{Lotz_2008}. \cite{French_2020} analyzed host galaxies of TDEs and quantified the asymmetry of their stellar light in the residual image. We follow this approach and use the residuals of the \texttt{GALFIT} model for the F625W band (see discussion in \S\ref{subsec: hst} and the top panel of Fig.~\ref{fig: HST_GALFIT_images} in Appendix \ref{sec: plots_appendix}) while masking the transient and the nearby galaxy. We find a low asymmetry measure of $A \simeq 0.04$ which implies that there was no recent major merger. However, this analysis does not exclude a minor merger, whose signatures may be faint.

\begin{figure}[ht]
\centering
\includegraphics[width=\linewidth]{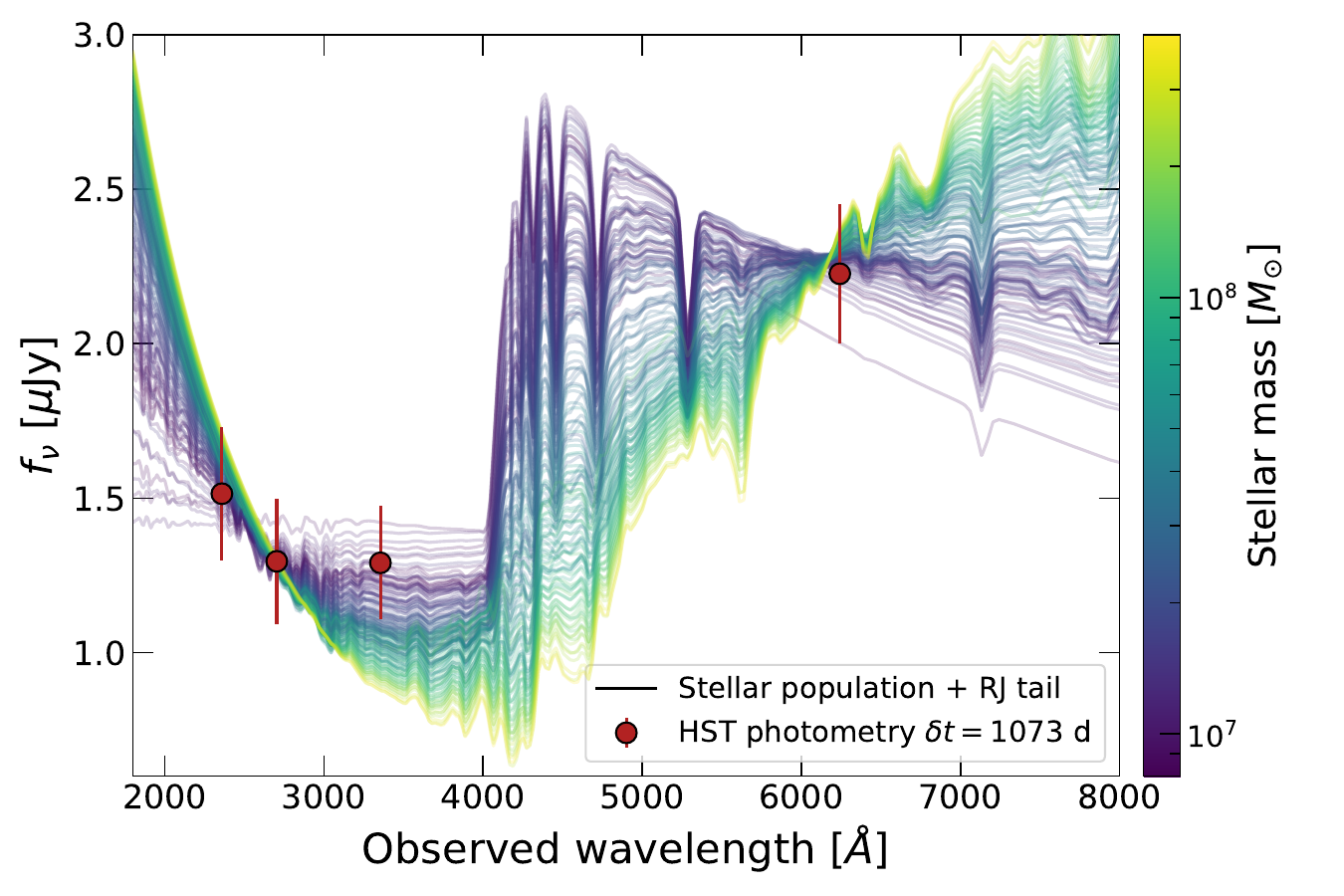}
\caption{HST flux measurements obtained on $\delta t = 1073$\,d (red circles; see \S\ref{subsec: hst} for details). These observations were fitted with a constant times RJ tail plus a constant times a stellar population component. We used \texttt{fsps} to generate a stellar population from a grid of stellar ages ($0.1-10$\,Gyr) and $\log Z$ ($-2.0 - 1.0$), and fitted the constants for each combination of age and $\log Z$. The best-fitting profiles are plotted in lines and color-coded by the stellar mass. The range of stellar mass we infer is $\sim 10^{7}-10^{8} \, M_{\odot}$.
\label{fig: late_time_phot}}
\end{figure}

In addition to the analysis of the morphology of the host, we also search for signs of an underlying stellar population at the position of the off-nuclear transient. This can be indicative of the an underlying nuclear star cluster (NSC; e.g., from a stripped host of the off-nuclear transient) or a background, or foreground, dwarf galaxy. With a typical size of an NSC $\sim 5$\,pc \citep{Neumayer_2020}, at the distance of TDE\,2023mfm, the angular size of an NSC will be $\sim 2.5$\,mas, and therefore expected to be detected as a point source. Our photometric measurements of the transient at late times ($\delta t \simeq 1073$\,d) with HST show an excess of emission in the optical band (F625W) compared to the other three bands. In Fig.~\ref{fig: late_time_phot} we show that the optical-to-UV SED can be described by a combination of a Rayleigh-Jeans (RJ) tail of a blackbody for the emission from the transient itself (as the late-time optical/UV emission of TDEs is expected to be dominated by thermal emission from the accretion disk; \citealt{Mummery_2020}), and a stellar population component that accounts for the excess in the optical wavelengths. 

Here we explored a simple stellar population (using the python implementation of \texttt{fsps}; \citealt{benjamin_johnson_2026_21778582}) and used a grid for the stellar population age (in the range of $0.1-10$\,Gyr) and for $\log Z$ (in the range of $-2.0 - 1.0$), and fitted for the constants scaling the RJ tail component and the stellar population component. Based on this analysis we find that if the excess seen in the late-time optical emission is due to a stellar component, the age of the stellar population is consistent with all the values we explored, and its mass is in the range of $\sim 10^{7}-10^{8} \, M_{\odot}$ (see Fig.~\ref{fig: late_time_phot}). Finally, to better constrain the shape of the SED, and therefore the age and mass of this stellar component, more HST and James Webb Space Telescope (JWST) observations are needed.

\subsubsection{Stellar population and kinematics}
\label{subsec: stellar_pop}

Pre-flare images of the host of TDE\,2023mfm from SDSS, \emph{Swift} UVOT, and GALEX NUV and FUV were used to construct a pre-flare photometric spectral energy distribution (SED). We then used \texttt{Prospector} software \citep{Johnson_2021} to run a Markov Chain Monte Carlo sampler \citep{Foreman_Mackey_2013}, with 100 walkers and 20,000 steps (discarding the first 2,000 steps as burn-in), to obtain the posterior distributions of the Flexible Stellar Population Synthesis models \citep{Conroy_2009}. The fitted parameters were the stellar mass, $M_{\rm gal}$, \citet{Calzetti_2000} dust model optical depth, stellar population age, $t_{\rm age}$, log metallicity, $\log Z$, and the e-folding time of the star formation history, $\tau_{\rm SFH}$. The chain
length exceeded $50\tau_{\rm corr}$ and effective sample sizes exceeded $\sim$400 for all five free parameters. We additionally compute $\hat{R}$, finding $\hat{R} \approx 1.04$ across all five parameters. We find $\log_{10} \left( M_{*} / M_{\odot} \right)= 10.95^{+0.13} _{-0.08}$, $\log Z = -0.64^{+0.29} _{-0.36}$, $t_{\rm age} = 9.2^{+3.7} _{-2.4}$\,Gyr, $\tau_{\rm SFH} = 1.5^{+0.6} _{-0.4}$, and $E \left( B-V\right)=0.6 \pm 0.2$. Based on the $M_{*}-M_{\rm BH}$ relations from \cite{Greene_2020} for early type galaxies we find that the central black hole has a mass of $M_{\rm BH} = 10^{8.53 \pm 0.80} \, M_{\odot}$ (where we include the intrinsic scatter of $0.65$\,dex). As discussed in \S\ref{subsec: swift_uv}, we used this \texttt{Prospector} fit to estimate the emission from the host and subtract it from the Swift/UVOT photometry.

Adding to this analysis of the photometric SED of the host, we also use the \texttt{pPXF} package \citep{Cappellari_2023} to fit the spectrum of the host obtained with LRIS with a stellar continuum model utilizing the \texttt{E-MILES} stellar population synthesis templates \citep{Vazdekis_2016}. From this fit we estimate a velocity dispersion measure of $\sigma_{\star} = 181.5 \pm 8.3 \, \rm km \, s^{-1}$. Assuming the $M_{\rm BH}-\sigma$ relations from \cite{Kormendy_2013} and an intrinsic scatter of $0.29$\,dex we estimate that the mass of the central black hole is $M_{\rm BH} = 10^{8.31 \pm 0.30} \, M_{\odot}$, consistent with the mass inferred from the \texttt{Prospector} fit to the pre-flare SED of the galaxy. 

\subsubsection{A low-luminosity AGN at the center}
\label{subsubsec: low_lum_agn}

As described in \S\ref{subsec: vla} and presented in Fig.~\ref{fig: optical_radio_images}, there are two point sources around the galactic nucleus in our radio observations. One is consistent with the center of the host galaxy seen in the HST images. The luminosity of this central source in the $10$\,GHz band is $\nu L_{\rm \nu} = \left( 9.0 \pm 1.2 \right) \times 10^{37} \, \rm erg \, s^{-1}$ and the spectral shape is consistent with a flat spectrum (see Fig.~\ref{fig: host_radio_sed} in Appendix~\ref{sec: plots_appendix}). There are two possible scenarios for the origin of the emission from the center of the host - emission from an active galactic nucleus (AGN) or star formation at the center. In \S\ref{subsec: chandra} we argue that the X-ray luminosity and spectral shape of the central source are consistent with a low-luminosity AGN. Here we use the optical and radio observations to test nuclear star formation versus a central MBH responsible for a low-luminosity AGN.

We first test both scenarios by examining the optical spectrum of the host. We subtract the best-fitting \texttt{pPXF} model (see \S\ref{subsec: stellar_pop}) and fit the remaining emission lines with Gaussian profiles. The H$\alpha$ and [\ion{N}{2}] lines are fitted by three Gaussian components and a constant continuum, the [\ion{S}{2}] doublet is fitted by two Gaussian components and a constant continuum, and the H$\beta$, [\ion{O}{1}] and [\ion{O}{3}] emission lines are each fitted by one Gaussian and a constant continuum. We find the following line ratios: $\log_{10}$([\ion{O}{3}/H$\beta$])~$=1.95 \pm 1.25$, $\log_{10}$([\ion{N}{2}/H$\alpha$])~$= 1.21 \pm 0.17$, $\log_{10}$([\ion{S}{2}/H$\alpha$])~$= 0.63 \pm 0.16$, and $\log_{10}$([\ion{O}{1}/H$\alpha$])~$= 0.31 \pm 0.09$. These ratios are consist with a low-ionization nuclear emission-line region (LINER) rather than with an \ion{H}{2} region or a Seyfert \citep{BPT_paper, Kewley_2006}. We estimate from this fit that the H$\alpha$ luminosity is $\left( 4.5 \pm 0.5 \right) \times 10^{39} \rm \, erg \, s^{-1}$. If we assume that all of this emission originates from star formation at the center, and use the scaling of the H$\alpha$ luminosity with star formation rate (SFR) introduced in \cite{Gallego_1995}, we estimate SFR\,$\leq 0.05 \, \rm{M_{\odot} \, yr^{-1}}$.

The radio emission from the central source is not consistent with the typical, optically thin, $F_{\rm \nu} \sim \nu^{-0.8}$, emission observed from star forming galaxies \citep{Magnelli_2015}. On the other hand, while the radio emission from AGN is often steep, $F_{\rm \nu} \sim \nu^{-0.7}$, and associated with optically thin synchrotron emission from extended radio lobes, flat radio SEDs have been observed for AGNs \citep{Tadhunter_2016, Kaiser_2006}. These are usually associated with multiple emission zones from relativistic jets (see e.g., the ``knot" ejection from the jet of a low-luminosity AGN; \citealt{King_2016}). Furthermore, assuming that the radio SED remains flat down to $1.4$\,GHz, and using the relation between the SFR and the luminosity in $1.4$\,GHz from \cite{Davies_2017} we infer an SFR of $\sim 0.6 \, \rm M_{\odot} \, yr^{-1}$, at least an order of magnitude higher than the upper limit on the SFR inferred from the H$\alpha$ luminosity. 

Therefore, overall, the emission from the center of the host is more consistent with a low-luminosity AGN than with star formation, and we conclude that the host of TDE\,2023mfm hosts at least two MBHs at its center (the central low-luminosity AGN and the wandering MBH giving rise to TDE\,2023mfm).

\subsection{The Origin of the Off-Nuclear MBH}
\label{subsec: off_nuclear_analysis}

The offset of the MBH that triggered TDE\,2023mfm is seen from the ZTF alert astrometry (see Fig.~\ref{fig: positional_analysis_plot}), and confirmed by our high-resolution observations with the HST and the VLA (see Fig.~\ref{fig: optical_radio_images}). Based on our HST observations we find that the separation between the center of the host galaxy and the off-nuclear MBH giving rise to the TDE is $1.08 \pm 0.04$\,kpc (see \S\ref{subsec: hst}). However, the origin of this off-nuclear MBH is not clear. Due to the similarities between the TDE\,2024tvd and TDE\,2023mfm, in the following section we are following the discussion presented in \S4.3 of \cite{Yao_2025} for the origin of the off-nuclear MBH that bring about TDE\,2024tvd.

One possible astrophysical origin for the pair of MBHs at the galactic center is an in-spiraling pre-merger MBH system. In this scenario the offset MBH originates in a galaxy that went through a major or minor merger with the host of the central MBH. In \S\ref{subsec: galaxy_profile} we find, based on a small asymmetry measure ($A\sim0.04$), that it is unlikely that the host galaxy went through a a recent ($\lesssim 0.5$\,Gyr) major merger. However, we cannot exclude a minor merger, and we note the extended structure seen in the residual of the fit to the F625W image as a possible sign of minor merger activity. 

Furthermore, we find evidence for an underlying stellar population with mass of $10^7-10^8 \, M_{\odot}$ at the position of the TDE, possibly from the nuclear star cluster hosting this MBH (see Fig.~\ref{fig: late_time_phot} and the discussion in \S\ref{subsec: galaxy_profile}). It is possible that this stellar component is either a background (or foreground) low-mass dwarf galaxy, or an NSC hosting this MBH. In the merger scenario, a significant time delay between the minor merger and the merger of the two MBHs is expected both from dynamical friction calculations \citep{Dosopoulou_2017} and cosmological simulations \citep{Tremmel_2018_pair}. Therefore, we conclude that it is possible that the offset MBH that produce the TDE is an in-spiraling MBH following a minor merger. Since the MBH is too massive ($10^6 - 10^7 \, M_{\odot}$) compared to the stellar mass at its position ($10^7 - 10^8 \, M_{\odot}$), we conclude that in this scenario, the merging galaxy has been stripped.

A different possible origin for this system is that the offset MBH was ejected from a triple MBH system due to gravitational wave recoil \citep{Komossa_2008, Stone_2012}. \cite{Yao_2025} estimated that the TDE rate from such recoiling MBHs is $\sim 1\%$ of the TDE rate. Since $\sim 250$ TDE candidates were identified in optical wavelengths so far \citep{Franz_2026}, it is possible that TDE\,2023mfm is from a recoiling MBH. Finally, another possibility is that the offset MBH got kicked by the GW emission following a merger of two MBHs. This will result in a single MBH outside of the galactic nucleus. The existence of an additional MBH confirmed by the low-luminosity AGN (see \S~\ref{subsec: host_analysis}) rules out both of these scenarios. 

Other than TDE\,2023mfm there are several off-nuclear TDEs and TDE candidates: 3XMM J2150 \citep{Lin_2018, Lin_2020}, EP240222a \citep{Jin_2025}, TDE\,2024tvd \citep{Yao_2025}, TDE\,2025abcr \citep{Stein_2026}, and EP250702a (\citealt{Levan_2025}; however, the TDE-nature of this candidate is still debated). All these candidates are similar in the total stellar mass of their parent galaxy ($\sim 10^{11} \, M_{\odot}$). On the other hand, the offset of 3XMM J2150, EP240222a, EP250702a, and TDE\,2025abcr from the center of their host is significantly larger ($> 5$\,kpc) than the offset of TDE\,2023mfm and TDE\,2024tvd ($\sim 1$\,kpc). This, together with the observational similarities between TDE\,2023mfm and TDE\,2024tvd (such as $t_{1/2}$, $L_{\rm g, \, peak}$, and $M_{\rm BH}$), is suggestive of a similar astrophysical origin to the MBHs of both TDE\,2023mfm and TDE\,2024tvd. On the other hand, we note that unlike TDE\,2023mfm, TDE\,2024tvd exhibit bright emission in both the X-ray and radio wavelengths at late-time. We speculate on the possible origin for this difference in the next section (\S\ref{subsec: radio_outflow_analysis}).

\subsection{The Nature of the Radio Emitting Outflow}
\label{subsec: radio_outflow_analysis}

\begin{figure}[ht]
\centering
\includegraphics[width=\linewidth]{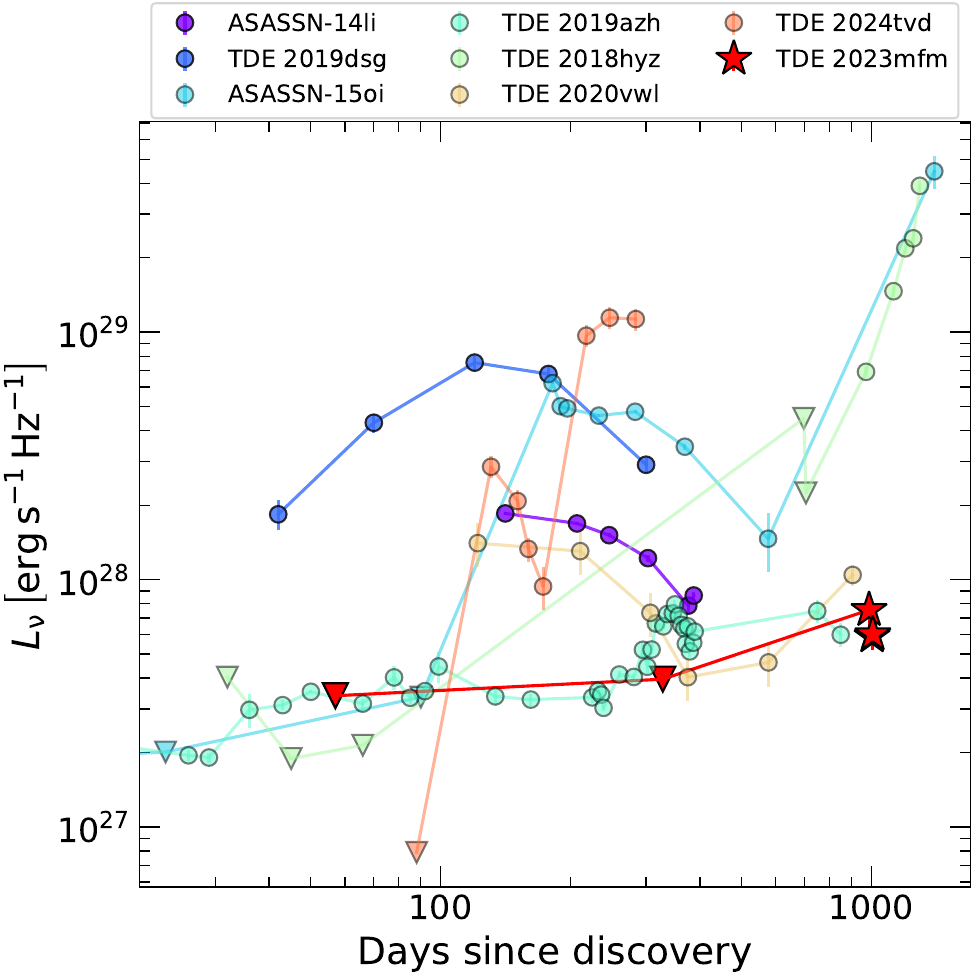}
\caption{Radio luminosities of selected radio-detected TDEs, and of TDE\,2023mfm (in red). Circles and stars mark detections and triangles mark $3\sigma$ upper limits. The radio emission from TDE\,2023mfm is faint compared to most radio detected TDEs.
\label{fig: radio_lc_comparison}}
\end{figure}

TDE\,2023mfm is not detected in the radio on $\delta t \simeq 60$ and $331$\,d (at $15$ and $6$\,GHz, respectively) and is first detected as a faint, $L_{\nu} \sim 7 \times 10^{27} \, \rm erg \, s^{-1} \, Hz^{-1}$, on $\delta t \simeq 991$\,d (see \S\ref{subsec: vla} and Fig.~\ref{fig: radio_lc_comparison}). The broadband emission ($3-15$\,GHz) is optically thin, and consistent with $F_{\rm \nu} \sim \nu^{-0.75}$ (see Fig.~\ref{fig: TDE_radio_emission} in Appendix \ref{sec: plots_appendix}). In comparison to the optically selected off-nuclear TDEs, TDE\,2024tvd had two delayed radio flares (peaking at $L_{\nu} \sim 3 \times 10^{28}  \, \rm erg \, s^{-1} Hz^{-1}$ during the first flare and $\sim 10^{29} \, \rm erg \, s^{-1} Hz^{-1}$ during the second flare) which were associated with at least one delayed outflow \citep{Sfaradi_2025}. TDE\,2025abcr, on the other hand, was not detected in the radio with an upper limit of $L_{\rm \nu} < 8 \times 10^{26} \, \rm erg \, s^{-1} \, Hz^{-1}$ \citep{Stein_2026}. Compared to the other off-nuclear TDE-candidates, no radio emission was detected for 3XMM J2150 and EP240222a. The bright radio emission from EP250702a ($L_{\nu} \sim 3 \times 10^{31} \, \rm erg \, s^{-1} \, Hz^{-1}$) is associated with a relativistic jet \citep{Levan_2025, Goodwin_2026}, however, the TDE-nature of EP250702a is still under debate.

Our radio observations of TDE\,2023mfm do not allow us to conduct the detailed analysis needed to determine the properties of the outflows and its origin. Based on the broadband SED reconstructed from observations obtained from $\delta t = 991$ to $1024$\,d, and following the non-relativistic formalism presented in \cite{Sfaradi_2025}, we estimate that radius of the outflow is $R \geq 3.1 \times 10^{16} \, \rm cm$, and that the post-shock energy is $U_{\rm ps} \geq 1.3 \times 10^{48} \, \rm erg$. This assumes equipartition, $\epsilon_{\rm e} = \epsilon_{\rm B} = 0.1$, where $\epsilon_{\rm e}$ is the fraction of energy deposited in the relativistic electrons and $\epsilon_{\rm B}$ is the fraction of energy deposited in the magnetic fields, an emission filling factor of $f=0.5$, and a power-law index for the electrons $p=2.5$, which is consistent with the observed spectral slope (assuming $F_{\nu} \sim \nu^{-(p-1)/2}$; \citealt{Sari_1998}). However, we emphasize that late-time brightening of TDEs in the radio might originate from an early launched off-axis relativistic jet \citep{Matsumoto_2023, Sfaradi_2024}. More observations, especially Very Long Baseline Interferometry (VLBI) observations, are needed to measure the size of the source and determine the nature of the radio emitting outflow from TDE\,2023mfm.

Based on the analysis of the transient optical/UV emission, and of the host, we concluded that TDE\,2023mfm shares observational similarities with the TDE\,2024tvd. The radio emission, on the other hand, tells a different story. As seen from Fig.~\ref{fig: radio_lc_comparison}, TDE\,2024tvd exhibits two fast-evolving, bright radio flares. The radio emission from TDE\,2023mfm, on the other hand, remains undetected around $\sim 300$ days (TDE\,2024tvd was radio-bright at these timescales), and is slowly evolving $\sim 3$ years after optical discovery. The difference between the radio properties of these two off-nuclear TDEs can be from (1) lack of a fast-moving outflow, (2) different CNM structure around the two different off-nuclear MBHs, and/or (3) viewing angle effects. Interestingly, for the TDE\,2024tvd, \citet{Sfaradi_2025} found a temporal correlation between the transition in the X-ray spectral state and the launch of the radio emitting outflow, hinting to an accretion-related delayed outflow producing the radio emission (also suggested for TDE\,2019azh; \citealt{Sfaradi_2022}). The lack of early X-ray emission from TDE\,2023mfm can be an indication that the difference between the radio properties of TDE\,2023mfm and TDE\,2024tvd is that an accretion-driven outflow was not formed early on (first $\sim$year).

\section{Conclusions}
\label{sec: conclusion}

We present the discovery of TDE\,2023mfm as an off-nuclear TDE at a distance of $1.08 \pm 0.04$\,kpc from the central MBH. This conclusion is supported by our radio, optical, and UV observations of TDE\,2023mfm. Specifically, our analysis of the optical and UV photometry as well as the optical spectroscopy clearly shows that TDE\,2023mfm is consistent with all the known properties of optically selected TDEs, and its classification as a TDE-H is based on the broad emission feature near H$\alpha$. Moreover, our combined analysis of our panchromatic (radio--optical--UV--X-rays) observations of the host reveal two MBHs near the galactic center. The central MBH is a low-luminosity AGN with a mass of $10^{8.31 \pm 0.30} \, M_{\odot}$. The off-nuclear MBH, with a mass of $10^{6.2 \pm 0.5} \, M_{\odot}$ (based on the UV plateau scaling relation), gave rise to the emission of TDE\,2023mfm.

TDE\,2023mfm shows all the defining observational characteristics of a TDE. The emission stays blue ($g-r \leq -0.2$\,mag) throughout its entire evolution and its slow rise and decline are incompatible with a supernova but match expectations for a TDE. Interestingly, when comparing to the other two optically selected off-nuclear TDEs, TDE\,2023mfm shows remarkable similarities to TDE\,2024tvd, unlike the clear difference between TDE\,2023mfm and TDE\,2025abcr. The time above half-maximum and the peak $g$-band luminosity for TDE\,2023mfm are almost identical to those of TDE\,2024tvd. The $g$-band emission from TDE\,2025abcr on the other hand rise faster to the peak, and is significantly fainter than TDE\,2023mfm. Furthermore, the physical offset between the TDE and the center of its host galaxy is $\sim 1$\,kpc for both TDE\,2023mfm and TDE\,2024tvd, unlike the other off-nuclear TDE candidates which are located at a larger distance ($\geq 5.5$\,kpc) from the center of their hosts. We note that these similarities might suggest a similar astrophysical origin for TDE\,2023mfm and TDE\,2024tvd. On the other hand, TDE\,2024tvd was brighter than TDE\,2023mfm is in both radio and X-rays, and showed a peculiar radio evolution. In \S\ref{subsec: radio_outflow_analysis} we speculate on this difference between TDE\,2023mfm and TDE\,2024tvd and consider either a lack of fast-moving outflow, a different CNM density around this off-nuclear MBH, or an off-axis jet. 

While the off-nuclear position of the MBH producing the TDE is certain, its astrophysical origin is unclear. Our analysis of the late-time optical/UV emission obtained with the HST rules out a recent major merger, but a recent minor merger is plausible. The minor merger scenario is extremely intriguing in the context of the evidence for an underlying stellar population with stellar mass of $\sim 10^7-10^8 \, M_{\odot}$ at the position of the off-nuclear MBH (see \S\ref{subsec: galaxy_profile}). Given the MBH mass ($10^6-10^7 \, M_{\odot}$), and the low mass of this stellar population compared to a full galaxy ($10^7-10^8 \, M_{\odot}$), we find that, in this minor merger scenario, the incoming galaxy has been significantly stripped.

Finally, we emphasize that high-resolution observations of TDEs and TDE-candidates are crucial to detect wandering MBHs. In particular, high-resolution radio observations provide a unique opportunity to prove the existence of more than one MBH at the center. If detected, the radio emission from the galactic center can be used to infer the existence of an AGN, and therefore a MBH at the center. If radio emission is also detected from the off-nuclear TDE this can provide a confirmation for the off-nuclear MBH and a precise measurement between the two MBHs. This was achieved in this work and for the TDE\,2024tvd \citep{Yao_2025, Sfaradi_2025}, and can be adopted by future searches for wandering MBHs.

\begin{acknowledgments}
Based on observations obtained with the Samuel Oschin Telescope 48-inch and the 60-inch Telescope at the Palomar Observatory as part of the Zwicky Transient Facility project. ZTF is supported by the National Science Foundation under Grants No. AST-2034437 and a collaboration including current partners Caltech, IPAC, the Oskar Klein Center at Stockholm University, the University of Maryland, University of California, Berkeley , the University of Wisconsin at Milwaukee, University of Warwick, Ruhr University Bochum, Cornell University, Northwestern University and Drexel University. Operations are conducted by COO, IPAC, and UW.
The ZTF forced-photometry service was funded under the Heising-Simons Foundation grant \#12540303 (PI: Graham).
The Gordon and Betty Moore Foundation, through both the Data-Driven Investigator Program and a dedicated grant, provided critical funding for SkyPortal.
SED Machine is based upon work supported by the National Science Foundation under Grant No. 1106171.
A major upgrade of the Kast spectrograph on the Shane 3\,m telescope at Lick Observatory, led by Brad Holden, was made possible through gifts from the Heising-Simons Foundation, William and Marina Kast, and the University of California Observatories. Research at Lick Observatory is partially supported by a generous gift from Google.
Some of the data presented herein were obtained at Keck Observatory, which is a private 501(c)3 nonprofit organization operated as a scientific partnership among the California Institute of Technology, the University of California, and the National Aeronautics and Space Administration. The Observatory was made possible by the generous financial support of the W. M. Keck Foundation. 
The authors wish to recognize and acknowledge the very significant cultural role and reverence that the summit of Maunakea has always had within the Native Hawaiian community. We are most fortunate to have the opportunity to conduct observations from this mountain.
Based in part on observations obtained with the NASA/ESA Hubble Space Telescope, retrieved from the Mikulski Archive for Space Telescopes (MAST) at the Space Telescope Science Institute (STScI). STScI is operated by the Association of Universities for Research in Astronomy, Inc. under NASA contract NAS 5-26555. Support for Program number GO-18131 was provided through a grant from the STScI under NASA contract NAS5-26555.
This research has made use of data obtained from the Chandra Data Archive provided by the Chandra X-ray Center (CXC).
Based in part on observations obtained with XMM-Newton, an ESA science mission with instruments and contributions directly funded by ESA Member States and NASA. R.C. acknowledges funding for the joint XMM-Newton/VLA AO-24 program \#96333 through NASA grant 80NSSC26K1056.
The National Radio Astronomy Observatory (NRAO) is a facility of the National Science Foundation operated under cooperative agreement by Associated Universities, Inc. We thank the NRAO for carrying out the Karl G. Jansky Very Large Array (VLA).

R.~M. acknowledges partial support from the National Science Foundation (grant number AST-2224255). 
H.S. acknowledges partial salary support from a Moore Foundation Postdoctoral Fellowship Grant to Rutgers University.
KDA and CTC acknowledge support provided by the NSF through award AST-2307668. KDA gratefully acknowledges support from the Alfred P. Sloan Foundation.

Portions of the spectral energy distribution fitting
code for the stellar population presented in Fig.~\ref{fig: late_time_phot} were developed with assistance from ChatGPT. All code was reviewed, tested, and validated by the authors.
\end{acknowledgments}

\facilities{HST (STIS), \emph{Swift} (UVOT), \emph{Swift} (XRT), Keck:I (LRIS), Shane (Kast), CXO (ACIS-S), XMM, VLA}

\software{
astropy \citep{2013A&A...558A..33A,2018AJ....156..123A},
scikit-learn \citep{scikit-learn},
CASA \citep{CASA},
\texttt{emcee} \citep{Foreman_Mackey_2013},
\texttt{pPXF} \citep{Cappellari_2023},
\texttt{Prospector} \citep{Johnson_2021}.
}

\appendix

\section{Data tables}
\label{sec: table_appendix}

In this section we attach the tables containing data and best-fitting parameters. Tables~\ref{tab: optical_photometry}, \ref{tab: swift_uvot_photometry}, and \ref{tab: hst_photometry}, present the optical and UV photometry (after host subtraction and correction from Galactic extinction) obtained with the ZTF, \emph{Swift} UVOT, and HST, respectively. Tables~\ref{tab: tde_radio_observation} and \ref{tab: host_radio_observation} present the flux measurements in radio wavelengths at the position of the transient and the center of the host, respectively.

\begin{deluxetable}{cccc}
\tablewidth{\columnwidth}
\tablecaption{ZTF optical photometry of TDE\,2023mfm.}
\tablehead{
\colhead{MJD} & \colhead{$\delta t$} & \colhead{Filter} &
\colhead{m} \\
\colhead{} & \colhead{$\rm \left[ Days \right]$} & \colhead{} & \colhead{$\rm \left[ AB \,mag \right]$}}
\startdata
60116.4 & -33.8 & i & $ > 20.08$ \\
60117.4 & -32.8 & r & $ > 20.69$ \\
60117.4 & -32.8 & g & $ > 20.66$ \\
60119.4 & -30.9 & i & $ > 20.13$ \\
60119.4 & -30.8 & g & $ > 20.75$ \\
60119.4 & -30.8 & r & $ > 20.91$ \\
60121.4 & -28.8 & g & $20.39 \pm 0.19$ \\
60121.4 & -28.8 & r & $ > 20.59$ \\
60122.4 & -27.8 & i & $ > 20.32$ \\
60123.4 & -26.8 & g & $19.8 \pm 0.14$ \\
60123.4 & -26.8 & r & $20.38 \pm 0.18$ \\
\enddata
\tablecomments{Summary of the optical photometry obtained with the ZTF for TDE\,2023mfm. $\delta t$ is the time in days since the peak of the $g$ band light curve and $m$ is the extinction-corrected AB magnitude. $5\sigma$ limits are reported when the source is not detected above the $5\sigma$ threshold. See \S\ref{subsec: ztf} for more details on the baseline subtraction and extinction-correction. A full version of this table is available online.
\label{tab: optical_photometry}}
\end{deluxetable}

\begin{deluxetable}{cccc}
\tablecaption{\emph{Swift} UVOT photometry of TDE\,2023mfm.}
\tablehead{
\colhead{MJD} & \colhead{$\delta t$} & \colhead{Filter} &
\colhead{m} \\
\colhead{} & \colhead{$\rm \left[ Days \right]$} & \colhead{} & \colhead{$\rm \left[ AB \,mag \right]$}}
\startdata
60166.43 & $16.2$ & \emph{uvm2} & $18.19 \pm 0.06$ \\
60166.43 & $16.2$ & \emph{uvw1} & $18.24 \pm 0.06$ \\
60166.43 & $16.2$ & \emph{uvw2} & $18.52 \pm 0.07$ \\
\enddata
\tablecomments{Summary of the UV photometry obtained with \emph{Swift} UVOT for TDE\,2023mfm. $\delta t$ is the time in days since the peak of the $g$ band light curve and $m$ is the extinction-corrected AB magnitude. 
\label{tab: swift_uvot_photometry}}
\end{deluxetable}

\begin{deluxetable}{cccc}
\tablecaption{HST photometry of TDE\,2023mfm.}
\tablehead{
\colhead{MJD} & \colhead{$\delta t$} & \colhead{Filter} &
\colhead{m} \\
\colhead{} & \colhead{$\rm \left[ Days \right]$} & \colhead{} & \colhead{$\rm \left[ AB \,mag \right]$}}
\startdata
61223 & $1073$ & F225W & $23.45 \pm 0.15$ \\
61223 & $1073$ & F275W & $23.62 \pm 0.17$ \\
61223 & $1073$ & F336W & $23.62 \pm 0.15$ \\
61223 & $1073$ & F625W & $23.03 \pm 0.11$ \\
\enddata
\tablecomments{Summary of the optical/UV photometry obtained with the HST for TDE\,2023mfm. $\delta t$ is the time in days since the peak of the $g$ band light curve and $m$ is the extinction-corrected AB magnitude. See \S\ref{subsec: hst} for more details on the point source fitting process.
\label{tab: hst_photometry}}
\end{deluxetable}

\begin{deluxetable*}{ccccccccc}[ht]
\tablecaption{VLA observations of TDE\,2023mfm.}
\tablehead{
\colhead{MJD} & \colhead{$\delta t$} & \colhead{Frequency} & \colhead{$F_{\rm \nu}$} & \colhead{Image rms} &
\colhead{Right Ascension} &
\colhead{Declination} &
\colhead{Program ID}\\
\colhead{} & \colhead{$\rm \left[ Days \right]$} & \colhead{$\rm \left[ GHz \right]$} & \colhead{$\rm \left[ mJy \right]$} & \colhead{$\rm \left[ mJy \right]$} & \colhead{} & \colhead{} & \colhead{}}
\startdata
$60210$ & $60$ & $15$ & $<0.018$ & $0.006$ & - & - & VLA 20B-377 \\
$60481$ & $331$ & $6$ & $<0.021$ & $0.007$ & - & - & VLA 20B-377 \\
$60481$ & $331$ & $10$ & $<0.018$ & $0.006$ & - & - & VLA 20B-377 \\
$61157$ & $1007$ & $3$ & $<0.051$ & $0.017$ & - & - & VLA 25B-109 \\
$61157$ & $1007$ & $6$ & $0.031 \pm 0.004$ & $0.006$ & 21:37:27.95638 & -04:20:43.18521 & VLA 25B-109 \\
$61157$ & $1007$ & $10$ & $<0.019$ & $0.0063$ & - & - & VLA 25B-109 \\
$61161$ & $1011$ & $6$ & $0.032 \pm 0.004$ & $0.003$ & 21:37:27.96010 & -04:20:43.13756 & VLA 25B-109 \\
$61174$ & $1024$ & $3$ & $0.046 \pm 0.005$ & $0.0065$ & 21:37:27.95807 & -04:20:43.10216 & XMM-Newton/VLA \#96333 \\
\enddata
\tablecomments{Summary of the radio observations obtained with the VLA for TDE\,2023mfm. $3\sigma$ upper limits are reported when the transient is not detected.
\label{tab: tde_radio_observation}}
\end{deluxetable*}

\begin{deluxetable*}{cccccccc}[ht]
\tablecaption{VLA observations of the point source at the center of the host.}
\tablehead{
\colhead{MJD} & \colhead{$\delta t$} & \colhead{Frequency} & \colhead{$F_{\rm \nu}$} & \colhead{Image rms} &
\colhead{Right Ascension} &
\colhead{Declination} &
\colhead{Program ID}\\
\colhead{} & \colhead{$\rm \left[ Days \right]$}& \colhead{$\rm \left[ GHz \right]$} & \colhead{$\rm \left[ mJy \right]$} & \colhead{$\rm \left[ mJy \right]$} & \colhead{} & \colhead{} & \colhead{}}
\startdata
$60210$ & $60$ & $15$ & $0.045 \pm 0.006$ & $0.006$ & 21:37:27.92477 & -04:20:42.70169 & VLA 20B-377 \\
$61157$ & $1007$ & $10$ & $0.052 \pm 0.007$ & $0.006$ & 21:37:27.92620 & -04:20:42.71719 & VLA 25B-109 \\
$61161$ & $1011$ & $6$ & $0.048 \pm 0.005$ & $0.003$ & 21:37:27.92469 & -04:20:42.71966 & VLA 25B-109 \\
$61174$ & $1024$ & $3$ & $0.032 \pm 0.004$ & $0.0065$ & 21:37:27.92938 & -04:20:42.69226 & XMM-Newton/VLA \#96333 \\
\enddata
\tablecomments{Summary of the radio observations obtained with the VLA for the source at the center of the host nucleus.
\label{tab: host_radio_observation}}
\end{deluxetable*}

\section{Plots}
\label{sec: plots_appendix}

\begin{figure*}[ht]
\centering
\includegraphics[width=\linewidth]{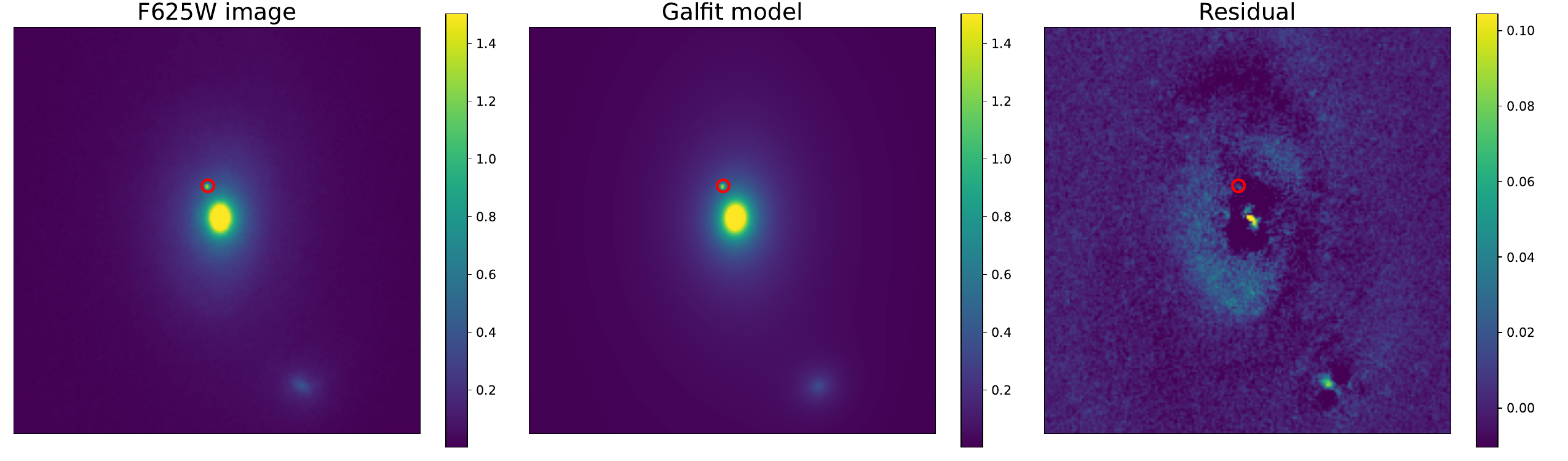}
\includegraphics[width=\linewidth]{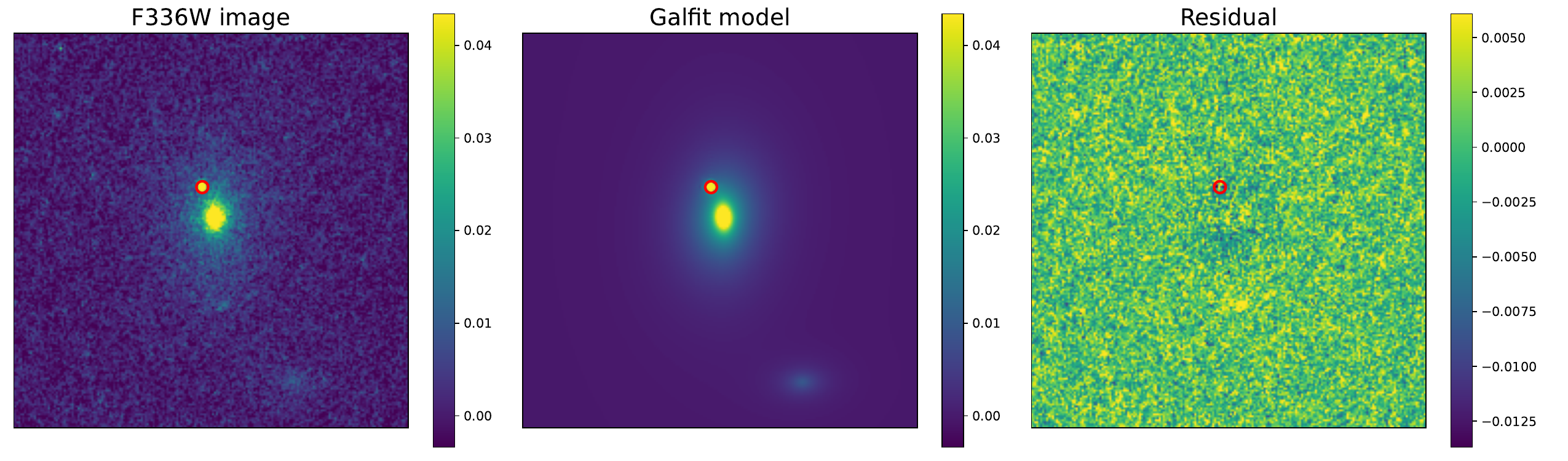}
\includegraphics[width=\linewidth]{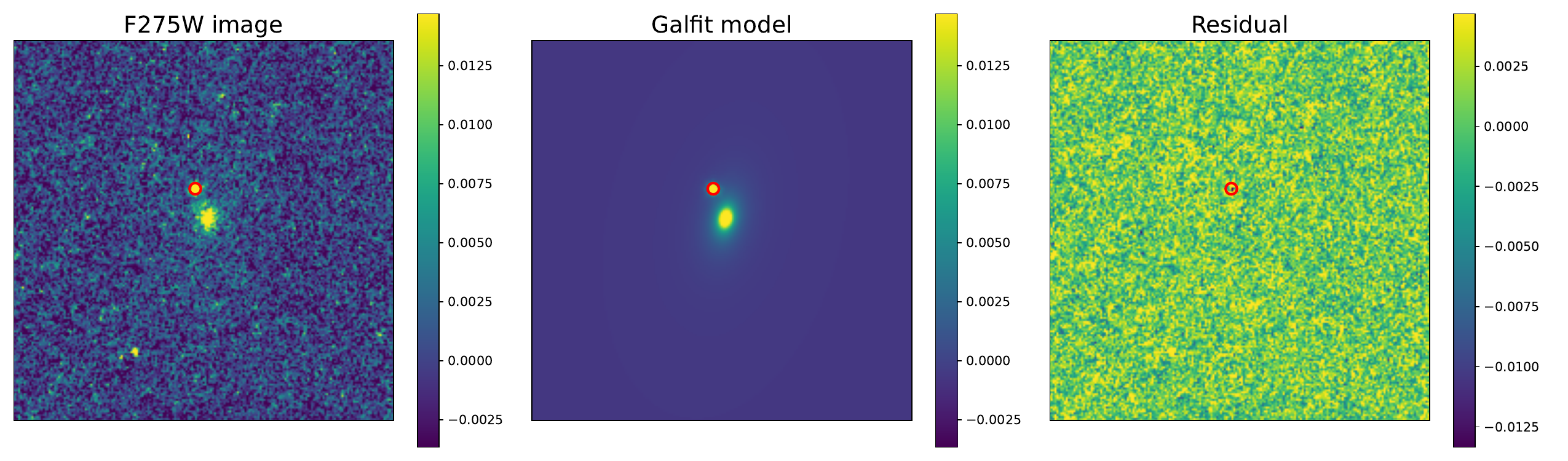}
\includegraphics[width=\linewidth]{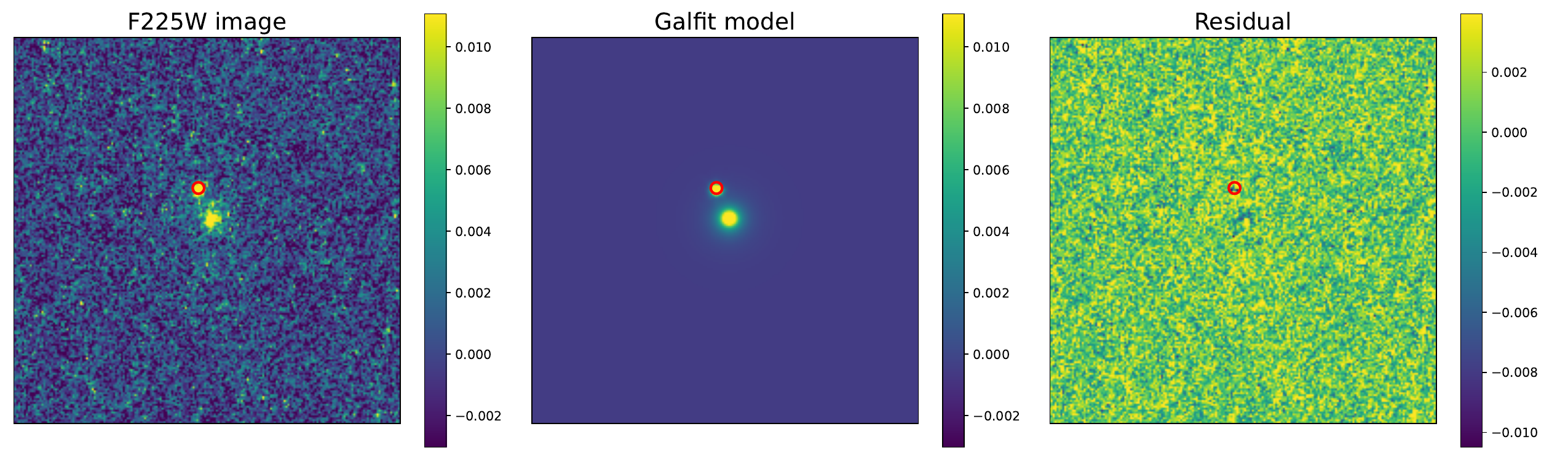}
\caption{HST science images (left panels), \texttt{GALFIT} models (middle panels), and residual images (right panels) of the TDE TDE\,2023mfm and its host. We mark the position of TDE\,2023mfm with a red circle. The first row of images is for F625W, the second row is for F336W, the third row is for F275W, and the fourth row is for F225W. See \S\ref{subsec: hst} for a detailed discussion on the fitting process.
\label{fig: HST_GALFIT_images}}
\end{figure*}

\begin{figure}[ht]
\centering
\includegraphics[width=\linewidth]{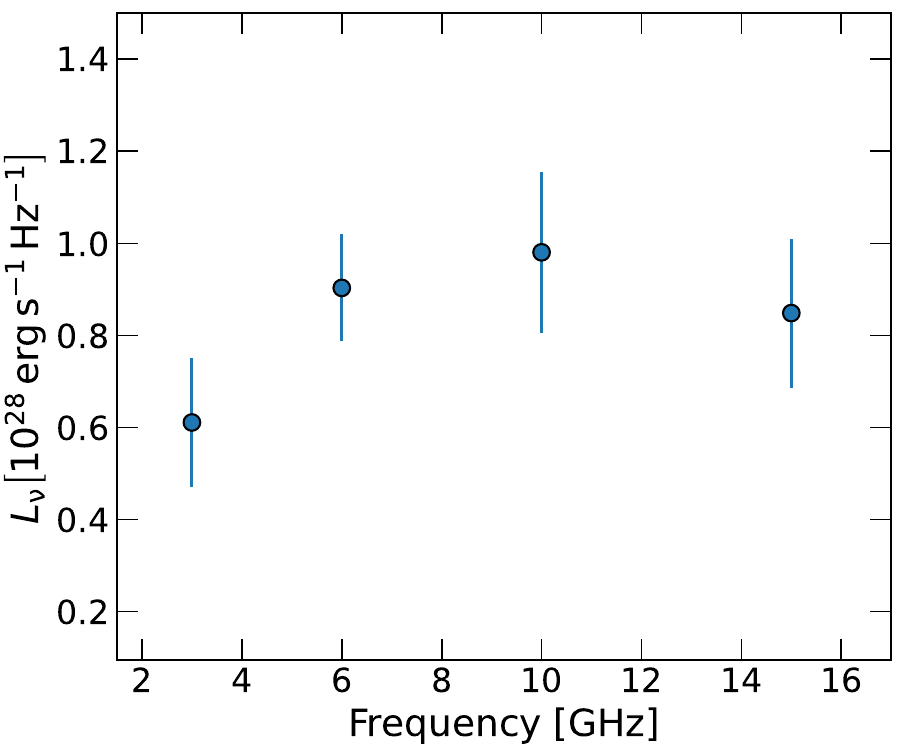}
\caption{Radio SED of the point source at the center of the host galaxy. The emission is consistent with a flat SED and we conclude that it is less likely to be from star formation, and more likely to be a low luminosity AGN (see the discussion in \S~\ref{subsec: host_analysis}).
\label{fig: host_radio_sed}}
\end{figure}

\begin{figure}[ht]
\centering
\includegraphics[width=\linewidth]{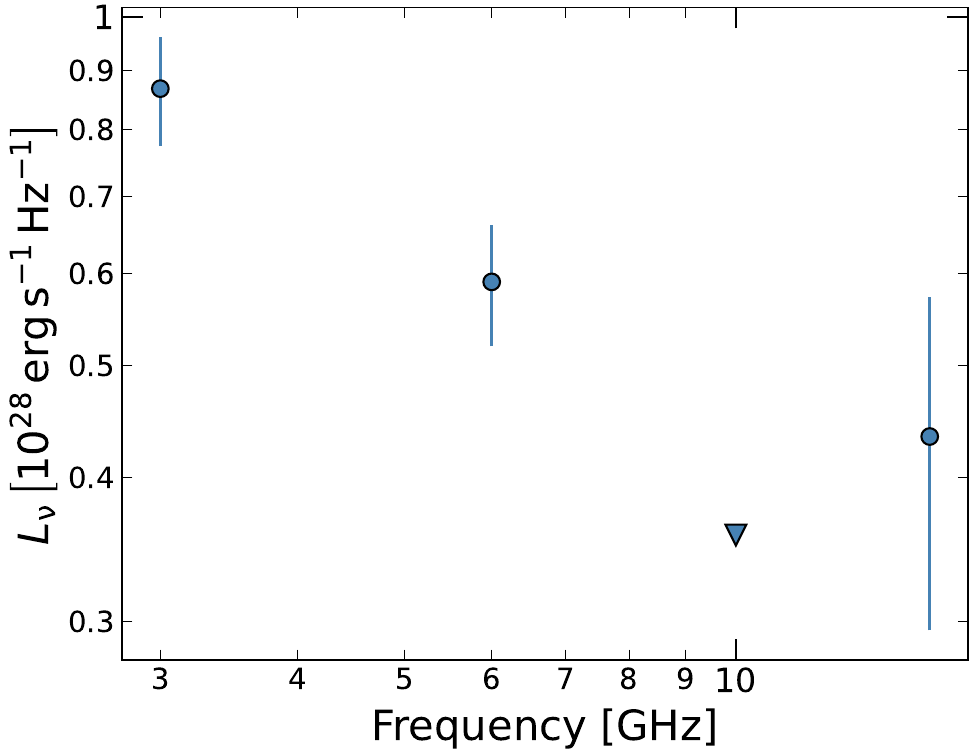}
\caption{Radio SED of the TDE\,2023mfm. The radio SED is optically thin down to $3$\,GHz (notice that the detection at $15$\,GHz is at $\gtrsim3\sigma$ image rms; \citealt{Wenkai_2026}).
\label{fig: TDE_radio_emission}}
\end{figure}

In this section we present supplementary plots. In Fig.~\ref{fig: HST_GALFIT_images} we show the HST images, the corresponding \texttt{GALFIT} models, and the residual images. Figures \ref{fig: host_radio_sed} and \ref{fig: TDE_radio_emission} present the radio SEDs of the point source at the galactic center and at the position of TDE\,2023mfm, respectively.

\bibliography{sample701}{}

\begin{thebibliography}{}
\expandafter\ifx\csname natexlab\endcsname\relax\def\natexlab#1{#1}\fi
\providecommand{\url}[1]{\href{#1}{#1}}
\providecommand{\dodoi}[1]{doi:~\href{http://doi.org/#1}{\nolinkurl{#1}}}
\providecommand{\doeprint}[1]{\href{http://ascl.net/#1}{\nolinkurl{http://ascl.net/#1}}}
\providecommand{\doarXiv}[1]{\href{https://arxiv.org/abs/#1}{\nolinkurl{https://arxiv.org/abs/#1}}}

\bibitem[{K.~D. {Alexander} {et~al.}(2020){Alexander}, {van Velzen}, {Horesh}, \& {Zauderer}}]{Alexander_2020}
{Alexander}, K.~D., {van Velzen}, S., {Horesh}, A., \& {Zauderer}, B.~A. 2020, \bibinfo{title}{{Radio Properties of Tidal Disruption Events},} \ssr, 216, 81, \dodoi{10.1007/s11214-020-00702-w}

\bibitem[{ {Astropy Collaboration} {et~al.}(2013){Astropy Collaboration}, {Robitaille}, {Tollerud}, {Greenfield}, {Droettboom}, {Bray}, {Aldcroft}, {Davis}, {Ginsburg}, {Price-Whelan}, {Kerzendorf}, {Conley}, {Crighton}, {Barbary}, {Muna}, {Ferguson}, {Grollier}, {Parikh}, {Nair}, {Unther}, {Deil}, {Woillez}, {Conseil}, {Kramer}, {Turner}, {Singer}, {Fox}, {Weaver}, {Zabalza}, {Edwards}, {Azalee Bostroem}, {Burke}, {Casey}, {Crawford}, {Dencheva}, {Ely}, {Jenness}, {Labrie}, {Lim}, {Pierfederici}, {Pontzen}, {Ptak}, {Refsdal}, {Servillat}, \& {Streicher}}]{2013A&A...558A..33A}
{Astropy Collaboration}, {Robitaille}, T.~P., {Tollerud}, E.~J., {et~al.} 2013, \bibinfo{title}{{Astropy: A community Python package for astronomy},} \aap, 558, A33, \dodoi{10.1051/0004-6361/201322068}

\bibitem[{ {Astropy Collaboration} {et~al.}(2018){Astropy Collaboration}, {Price-Whelan}, {Sip{\H{o}}cz}, {G{\"u}nther}, {Lim}, {Crawford}, {Conseil}, {Shupe}, {Craig}, {Dencheva}, {Ginsburg}, {VanderPlas}, {Bradley}, {P{\'e}rez-Su{\'a}rez}, {de Val-Borro}, {Aldcroft}, {Cruz}, {Robitaille}, {Tollerud}, {Ardelean}, {Babej}, {Bach}, {Bachetti}, {Bakanov}, {Bamford}, {Barentsen}, {Barmby}, {Baumbach}, {Berry}, {Biscani}, {Boquien}, {Bostroem}, {Bouma}, {Brammer}, {Bray}, {Breytenbach}, {Buddelmeijer}, {Burke}, {Calderone}, {Cano Rodr{\'\i}guez}, {Cara}, {Cardoso}, {Cheedella}, {Copin}, {Corrales}, {Crichton}, {D'Avella}, {Deil}, {Depagne}, {Dietrich}, {Donath}, {Droettboom}, {Earl}, {Erben}, {Fabbro}, {Ferreira}, {Finethy}, {Fox}, {Garrison}, {Gibbons}, {Goldstein}, {Gommers}, {Greco}, {Greenfield}, {Groener}, {Grollier}, {Hagen}, {Hirst}, {Homeier}, {Horton}, {Hosseinzadeh}, {Hu}, {Hunkeler}, {Ivezi{\'c}}, {Jain}, {Jenness}, {Kanarek}, {Kendrew}, {Kern}, {Kerzendorf}, {Khvalko}, {King}, {Kirkby}, {Kulkarni},
  {Kumar}, {Lee}, {Lenz}, {Littlefair}, {Ma}, {Macleod}, {Mastropietro}, {McCully}, {Montagnac}, {Morris}, {Mueller}, {Mumford}, {Muna}, {Murphy}, {Nelson}, {Nguyen}, {Ninan}, {N{\"o}the}, {Ogaz}, {Oh}, {Parejko}, {Parley}, {Pascual}, {Patil}, {Patil}, {Plunkett}, {Prochaska}, {Rastogi}, {Reddy Janga}, {Sabater}, {Sakurikar}, {Seifert}, {Sherbert}, {Sherwood-Taylor}, {Shih}, {Sick}, {Silbiger}, {Singanamalla}, {Singer}, {Sladen}, {Sooley}, {Sornarajah}, {Streicher}, {Teuben}, {Thomas}, {Tremblay}, {Turner}, {Terr{\'o}n}, {van Kerkwijk}, {de la Vega}, {Watkins}, {Weaver}, {Whitmore}, {Woillez}, {Zabalza}, \& {Astropy Contributors}}]{2018AJ....156..123A}
{Astropy Collaboration}, {Price-Whelan}, A.~M., {Sip{\H{o}}cz}, B.~M., {et~al.} 2018, \bibinfo{title}{{The Astropy Project: Building an Open-science Project and Status of the v2.0 Core Package},} \aj, 156, 123, \dodoi{10.3847/1538-3881/aabc4f}

\bibitem[{J.~A. {Baldwin} {et~al.}(1981){Baldwin}, {Phillips}, \& {Terlevich}}]{BPT_paper}
{Baldwin}, J.~A., {Phillips}, M.~M., \& {Terlevich}, R. 1981, \bibinfo{title}{{Classification parameters for the emission-line spectra of extragalactic objects.},} \pasp, 93, 5, \dodoi{10.1086/130766}

\bibitem[{J.~D. {Bekenstein}(1973){Bekenstein}}]{Bekenstein_1973}
{Bekenstein}, J.~D. 1973, \bibinfo{title}{{Gravitational-Radiation Recoil and Runaway Black Holes},} \apj, 183, 657, \dodoi{10.1086/152255}

\bibitem[{E.~C. {Bellm} {et~al.}(2019){Bellm}, {Kulkarni}, {Graham}, {Dekany}, {Smith}, {Riddle}, {Masci}, {Helou}, {Prince}, {Adams}, {Barbarino}, {Barlow}, {Bauer}, {Beck}, {Belicki}, {Biswas}, {Blagorodnova}, {Bodewits}, {Bolin}, {Brinnel}, {Brooke}, {Bue}, {Bulla}, {Burruss}, {Cenko}, {Chang}, {Connolly}, {Coughlin}, {Cromer}, {Cunningham}, {De}, {Delacroix}, {Desai}, {Duev}, {Eadie}, {Farnham}, {Feeney}, {Feindt}, {Flynn}, {Franckowiak}, {Frederick}, {Fremling}, {Gal-Yam}, {Gezari}, {Giomi}, {Goldstein}, {Golkhou}, {Goobar}, {Groom}, {Hacopians}, {Hale}, {Henning}, {Ho}, {Hover}, {Howell}, {Hung}, {Huppenkothen}, {Imel}, {Ip}, {Ivezi{\'c}}, {Jackson}, {Jones}, {Juric}, {Kasliwal}, {Kaspi}, {Kaye}, {Kelley}, {Kowalski}, {Kramer}, {Kupfer}, {Landry}, {Laher}, {Lee}, {Lin}, {Lin}, {Lunnan}, {Giomi}, {Mahabal}, {Mao}, {Miller}, {Monkewitz}, {Murphy}, {Ngeow}, {Nordin}, {Nugent}, {Ofek}, {Patterson}, {Penprase}, {Porter}, {Rauch}, {Rebbapragada}, {Reiley}, {Rigault}, {Rodriguez}, {van Roestel}, {Rusholme},
  {van Santen}, {Schulze}, {Shupe}, {Singer}, {Soumagnac}, {Stein}, {Surace}, {Sollerman}, {Szkody}, {Taddia}, {Terek}, {Van Sistine}, {van Velzen}, {Vestrand}, {Walters}, {Ward}, {Ye}, {Yu}, {Yan}, \& {Zolkower}}]{Bellm_2019}
{Bellm}, E.~C., {Kulkarni}, S.~R., {Graham}, M.~J., {et~al.} 2019, \bibinfo{title}{{The Zwicky Transient Facility: System Overview, Performance, and First Results},} \pasp, 131, 018002, \dodoi{10.1088/1538-3873/aaecbe}

\bibitem[{N. {Blagorodnova} {et~al.}(2018){Blagorodnova}, {Neill}, {Walters}, {Kulkarni}, {Fremling}, {Ben-Ami}, {Dekany}, {Fucik}, {Konidaris}, {Nash}, {Ngeow}, {Ofek}, {O' Sullivan}, {Quimby}, {Ritter}, \& {Vyhmeister}}]{Blagorodnova_2018}
{Blagorodnova}, N., {Neill}, J.~D., {Walters}, R., {et~al.} 2018, \bibinfo{title}{{The SED Machine: A Robotic Spectrograph for Fast Transient Classification},} \pasp, 130, 035003, \dodoi{10.1088/1538-3873/aaa53f}

\bibitem[{L. {Blecha} \& A. {Loeb}(2008){Blecha} \& {Loeb}}]{Blecha_2008}
{Blecha}, L., \& {Loeb}, A. 2008, \bibinfo{title}{{Effects of gravitational-wave recoil on the dynamics and growth of supermassive black holes},} \mnras, 390, 1311, \dodoi{10.1111/j.1365-2966.2008.13790.x}

\bibitem[{D. {Calzetti} {et~al.}(2000){Calzetti}, {Armus}, {Bohlin}, {Kinney}, {Koornneef}, \& {Storchi-Bergmann}}]{Calzetti_2000}
{Calzetti}, D., {Armus}, L., {Bohlin}, R.~C., {et~al.} 2000, \bibinfo{title}{{The Dust Content and Opacity of Actively Star-forming Galaxies},} \apj, 533, 682, \dodoi{10.1086/308692}

\bibitem[{M. {Cappellari}(2023){Cappellari}}]{Cappellari_2023}
{Cappellari}, M. 2023, \bibinfo{title}{{Full spectrum fitting with photometry in PPXF: stellar population versus dynamical masses, non-parametric star formation history and metallicity for 3200 LEGA-C galaxies at redshift z {\ensuremath{\approx}} 0.8},} \mnras, 526, 3273, \dodoi{10.1093/mnras/stad2597}

\bibitem[{J.~A. {Cardelli} {et~al.}(1989){Cardelli}, {Clayton}, \& {Mathis}}]{Cardelli_1989}
{Cardelli}, J.~A., {Clayton}, G.~C., \& {Mathis}, J.~S. 1989, \bibinfo{title}{{The Relationship between Infrared, Optical, and Ultraviolet Extinction},} \apj, 345, 245, \dodoi{10.1086/167900}

\bibitem[{C. {Carter} {et~al.}(2003){Carter}, {Karovska}, {Jerius}, {Glotfelty}, \& {Beikman}}]{Carter2003}
{Carter}, C., {Karovska}, M., {Jerius}, D., {Glotfelty}, K., \& {Beikman}, S. 2003, \bibinfo{title}{{ChaRT: The Chandra Ray Tracer},} in Astronomical Society of the Pacific Conference Series, Vol. 295, Astronomical Data Analysis Software and Systems XII, ed. H.~E. {Payne}, R.~I. {Jedrzejewski}, \& R.~N. {Hook}, 477

\bibitem[{ {CASA Team} {et~al.}(2022){CASA Team}, {Bean}, {Bhatnagar}, {Castro}, {Donovan Meyer}, {Emonts}, {Garcia}, {Garwood}, {Golap}, {Gonzalez Villalba}, {Harris}, {Hayashi}, {Hoskins}, {Hsieh}, {Jagannathan}, {Kawasaki}, {Keimpema}, {Kettenis}, {Lopez}, {Marvil}, {Masters}, {McNichols}, {Mehringer}, {Miel}, {Moellenbrock}, {Montesino}, {Nakazato}, {Ott}, {Petry}, {Pokorny}, {Raba}, {Rau}, {Schiebel}, {Schweighart}, {Sekhar}, {Shimada}, {Small}, {Steeb}, {Sugimoto}, {Suoranta}, {Tsutsumi}, {van Bemmel}, {Verkouter}, {Wells}, {Xiong}, {Szomoru}, {Griffith}, {Glendenning}, \& {Kern}}]{CASA}
{CASA Team}, {Bean}, B., {Bhatnagar}, S., {et~al.} 2022, \bibinfo{title}{{CASA, the Common Astronomy Software Applications for Radio Astronomy},} \pasp, 134, 114501, \dodoi{10.1088/1538-3873/ac9642}

\bibitem[{Y. {Cendes} {et~al.}(2024){Cendes}, {Berger}, {Alexander}, {Chornock}, {Margutti}, {Metzger}, {Wieringa}, {Bietenholz}, {Hajela}, {Laskar}, {Stroh}, \& {Terreran}}]{Cendes_2024}
{Cendes}, Y., {Berger}, E., {Alexander}, K.~D., {et~al.} 2024, \bibinfo{title}{{Ubiquitous Late Radio Emission from Tidal Disruption Events},} \apj, 971, 185, \dodoi{10.3847/1538-4357/ad5541}

\bibitem[{S.~B. {Cenko} {et~al.}(2006){Cenko}, {Fox}, {Moon}, {Harrison}, {Kulkarni}, {Henning}, {Guzman}, {Bonati}, {Smith}, {Thicksten}, {Doyle}, {Petrie}, {Gal-Yam}, {Soderberg}, {Anagnostou}, \& {Laity}}]{Cenko_2006}
{Cenko}, S.~B., {Fox}, D.~B., {Moon}, D.-S., {et~al.} 2006, \bibinfo{title}{{The Automated Palomar 60 Inch Telescope},} \pasp, 118, 1396, \dodoi{10.1086/508366}

\bibitem[{N. {Chen} {et~al.}(2023){Chen}, {Di Matteo}, {Ni}, {Tremmel}, {DeGraf}, {Shen}, {Holgado}, {Bird}, {Croft}, \& {Feng}}]{Chen_2023}
{Chen}, N., {Di Matteo}, T., {Ni}, Y., {et~al.} 2023, \bibinfo{title}{{Properties and evolution of dual and offset AGN in the ASTRID simulation at z 2},} \mnras, 522, 1895, \dodoi{10.1093/mnras/stad834}

\bibitem[{R. {Chornock}(2023){Chornock}}]{Chornock_2023_23mfm}
{Chornock}, R. 2023, \bibinfo{title}{{Transient Classification Report for 2023-09-18},} Transient Name Server Classification Report, 2023-2308, 1

\bibitem[{R. {Chornock} {et~al.}(2023){Chornock}, {Yao}, {LeBaron}, {Sears}, {Nicholl}, \& {Ridley}}]{Chornock_2023_23mfm_astronote}
{Chornock}, R., {Yao}, Y., {LeBaron}, N., {et~al.} 2023, \bibinfo{title}{{AT2023mfm: a Tidal Disruption Event at z=0.087},} Transient Name Server AstroNote, 250, 1

\bibitem[{J.~M. {Comerford} {et~al.}(2015){Comerford}, {Pooley}, {Barrows}, {Greene}, {Zakamska}, {Madejski}, \& {Cooper}}]{Comerford_2015}
{Comerford}, J.~M., {Pooley}, D., {Barrows}, R.~S., {et~al.} 2015, \bibinfo{title}{{Merger-driven Fueling of Active Galactic Nuclei: Six Dual and Offset AGNs Discovered with Chandra and Hubble Space Telescope Observations},} \apj, 806, 219, \dodoi{10.1088/0004-637X/806/2/219}

\bibitem[{C. {Conroy} \& R.~H. {Wechsler}(2009){Conroy} \& {Wechsler}}]{Conroy_2009}
{Conroy}, C., \& {Wechsler}, R.~H. 2009, \bibinfo{title}{{Connecting Galaxies, Halos, and Star Formation Rates Across Cosmic Time},} \apj, 696, 620, \dodoi{10.1088/0004-637X/696/1/620}

\bibitem[{M.~W. {Coughlin} {et~al.}(2023){Coughlin}, {Bloom}, {Nir}, {Antier}, {du Laz}, {van der Walt}, {Crellin-Quick}, {Culino}, {Duev}, {Goldstein}, {Healy}, {Karambelkar}, {Lilleboe}, {Shin}, {Singer}, {Ahumada}, {Anand}, {Bellm}, {Dekany}, {Graham}, {Kasliwal}, {Kostadinova}, {Kiendrebeogo}, {Kulkarni}, {Jenkins}, {LeBaron}, {Mahabal}, {Neill}, {Parazin}, {Peloton}, {Perley}, {Riddle}, {Rusholme}, {van Santen}, {Sollerman}, {Stein}, {Turpin}, {Wold}, {Amat}, {Bonnefon}, {Bonnefoy}, {Flament}, {Kerkow}, {Kishore}, {Jani}, {Mahanty}, {Liu}, {Llinares}, {Makarison}, {Olli{\'e}ric}, {Perez}, {Pont}, \& {Sharma}}]{Coughlin_2023}
{Coughlin}, M.~W., {Bloom}, J.~S., {Nir}, G., {et~al.} 2023, \bibinfo{title}{{A Data Science Platform to Enable Time-domain Astronomy},} \apjs, 267, 31, \dodoi{10.3847/1538-4365/acdee1}

\bibitem[{L.~J.~M. {Davies} {et~al.}(2017){Davies}, {Huynh}, {Hopkins}, {Seymour}, {Driver}, {Robotham}, {Baldry}, {Bland-Hawthorn}, {Bourne}, {Bremer}, {Brown}, {Brough}, {Cluver}, {Grootes}, {Jarvis}, {Loveday}, {Moffet}, {Owers}, {Phillipps}, {Sadler}, {Wang}, {Wilkins}, \& {Wright}}]{Davies_2017}
{Davies}, L.~J.~M., {Huynh}, M.~T., {Hopkins}, A.~M., {et~al.} 2017, \bibinfo{title}{{Galaxy And Mass Assembly: the 1.4 GHz SFR indicator, SFR-M$_{*}$ relation and predictions for ASKAP-GAMA},} \mnras, 466, 2312, \dodoi{10.1093/mnras/stw3080}

\bibitem[{J.~E. {Davis} {et~al.}(2012){Davis}, {Bautz}, {Dewey}, {Heilmann}, {Houck}, {Huenemoerder}, {Marshall}, {Nowak}, {Schattenburg}, {Schulz}, \& {Smith}}]{Davis2012}
{Davis}, J.~E., {Bautz}, M.~W., {Dewey}, D., {et~al.} 2012, \bibinfo{title}{{Raytracing with MARX: x-ray observatory design, calibration, and support},} in Society of Photo-Optical Instrumentation Engineers (SPIE) Conference Series, Vol. 8443, Space Telescopes and Instrumentation 2012: Ultraviolet to Gamma Ray, ed. T.~{Takahashi}, S.~S. {Murray}, \& J.-W.~A. {den Herder}, 84431A, \dodoi{10.1117/12.926937}

\bibitem[{A. {De Rosa} {et~al.}(2019){De Rosa}, {Vignali}, {Bogdanovi{\'c}}, {Capelo}, {Charisi}, {Dotti}, {Husemann}, {Lusso}, {Mayer}, {Paragi}, {Runnoe}, {Sesana}, {Steinborn}, {Bianchi}, {Colpi}, {del Valle}, {Frey}, {Gab{\'a}nyi}, {Giustini}, {Guainazzi}, {Haiman}, {Herrera Ruiz}, {Herrero-Illana}, {Iwasawa}, {Komossa}, {Lena}, {Loiseau}, {Perez-Torres}, {Piconcelli}, \& {Volonteri}}]{DeRosa_2019}
{De Rosa}, A., {Vignali}, C., {Bogdanovi{\'c}}, T., {et~al.} 2019, \bibinfo{title}{{The quest for dual and binary supermassive black holes: A multi-messenger view},} \nar, 86, 101525, \dodoi{10.1016/j.newar.2020.101525}

\bibitem[{R. {Dekany} {et~al.}(2020){Dekany}, {Smith}, {Riddle}, {Feeney}, {Porter}, {Hale}, {Zolkower}, {Belicki}, {Kaye}, {Henning}, {Walters}, {Cromer}, {Delacroix}, {Rodriguez}, {Reiley}, {Mao}, {Hover}, {Murphy}, {Burruss}, {Baker}, {Kowalski}, {Reif}, {Mueller}, {Bellm}, {Graham}, \& {Kulkarni}}]{Dekany_2020}
{Dekany}, R., {Smith}, R.~M., {Riddle}, R., {et~al.} 2020, \bibinfo{title}{{The Zwicky Transient Facility: Observing System},} \pasp, 132, 038001, \dodoi{10.1088/1538-3873/ab4ca2}

\bibitem[{F. {Dosopoulou} \& F. {Antonini}(2017){Dosopoulou} \& {Antonini}}]{Dosopoulou_2017}
{Dosopoulou}, F., \& {Antonini}, F. 2017, \bibinfo{title}{{Dynamical Friction and the Evolution of Supermassive Black Hole Binaries: The Final Hundred-parsec Problem},} \apj, 840, 31, \dodoi{10.3847/1538-4357/aa6b58}

\bibitem[{D. {Foreman-Mackey} {et~al.}(2013){Foreman-Mackey}, {Hogg}, {Lang}, \& {Goodman}}]{Foreman_Mackey_2013}
{Foreman-Mackey}, D., {Hogg}, D.~W., {Lang}, D., \& {Goodman}, J. 2013, \bibinfo{title}{{emcee: The MCMC Hammer},} \pasp, 125, 306, \dodoi{10.1086/670067}

\bibitem[{N. {Franz} {et~al.}(2026){Franz}, {Alexander}, {Gomez}, {Christy}, {Laskar}, {van Velzen}, {Earl}, {Gezari}, {Karmen}, {Margutti}, {Pearson}, {Villar}, \& {Zabludoff}}]{Franz_2026}
{Franz}, N., {Alexander}, K.~D., {Gomez}, S., {et~al.} 2026, \bibinfo{title}{{The Open mulTiwavelength Transient Event Repository (OTTER): Infrastructure Release and Tidal Disruption Event Catalog},} \apj, 999, 243, \dodoi{10.3847/1538-4357/ae346e}

\bibitem[{C. {Fremling}(2023){Fremling}}]{Fremling_2023_23mfm}
{Fremling}, C. 2023, \bibinfo{title}{{ZTF Transient Discovery Report for 2023-07-01},} Transient Name Server Discovery Report, 2023-1546, 1

\bibitem[{K.~D. {French} {et~al.}(2020){French}, {Wevers}, {Law-Smith}, {Graur}, \& {Zabludoff}}]{French_2020}
{French}, K.~D., {Wevers}, T., {Law-Smith}, J., {Graur}, O., \& {Zabludoff}, A.~I. 2020, \bibinfo{title}{{The Host Galaxies of Tidal Disruption Events},} \ssr, 216, 32, \dodoi{10.1007/s11214-020-00657-y}

\bibitem[{A. {Fruscione} {et~al.}(2006){Fruscione}, {McDowell}, {Allen}, {Brickhouse}, {Burke}, {Davis}, {Durham}, {Elvis}, {Galle}, {Harris}, {Huenemoerder}, {Houck}, {Ishibashi}, {Karovska}, {Nicastro}, {Noble}, {Nowak}, {Primini}, {Siemiginowska}, {Smith}, \& {Wise}}]{Fruscione2006}
{Fruscione}, A., {McDowell}, J.~C., {Allen}, G.~E., {et~al.} 2006, \bibinfo{title}{{CIAO: Chandra's data analysis system},} in Society of Photo-Optical Instrumentation Engineers (SPIE) Conference Series, Vol. 6270, Observatory Operations: Strategies, Processes, and Systems, ed. D.~R. {Silva} \& R.~E. {Doxsey}, 62701V, \dodoi{10.1117/12.671760}

\bibitem[{J. {Gallego} {et~al.}(1995){Gallego}, {Zamorano}, {Aragon-Salamanca}, \& {Rego}}]{Gallego_1995}
{Gallego}, J., {Zamorano}, J., {Aragon-Salamanca}, A., \& {Rego}, M. 1995, \bibinfo{title}{{The Current Star Formation Rate of the Local Universe},} \apjl, 455, L1, \dodoi{10.1086/309804}

\bibitem[{G.~P. {Garmire} {et~al.}(2003){Garmire}, {Bautz}, {Ford}, {Nousek}, \& {Ricker}}]{Garmire2003}
{Garmire}, G.~P., {Bautz}, M.~W., {Ford}, P.~G., {Nousek}, J.~A., \& {Ricker}, George~R., J. 2003, \bibinfo{title}{{Advanced CCD imaging spectrometer (ACIS) instrument on the Chandra X-ray Observatory},} in Society of Photo-Optical Instrumentation Engineers (SPIE) Conference Series, Vol. 4851, X-Ray and Gamma-Ray Telescopes and Instruments for Astronomy., ed. J.~E. {Truemper} \& H.~D. {Tananbaum}, 28--44, \dodoi{10.1117/12.461599}

\bibitem[{N. {Gehrels} {et~al.}(2004){Gehrels}, {Chincarini}, {Giommi}, {Mason}, {Nousek}, {Wells}, {White}, {Barthelmy}, {Burrows}, {Cominsky}, {Hurley}, {Marshall}, {M{\'e}sz{\'a}ros}, {Roming}, {Angelini}, {Barbier}, {Belloni}, {Campana}, {Caraveo}, {Chester}, {Citterio}, {Cline}, {Cropper}, {Cummings}, {Dean}, {Feigelson}, {Fenimore}, {Frail}, {Fruchter}, {Garmire}, {Gendreau}, {Ghisellini}, {Greiner}, {Hill}, {Hunsberger}, {Krimm}, {Kulkarni}, {Kumar}, {Lebrun}, {Lloyd-Ronning}, {Markwardt}, {Mattson}, {Mushotzky}, {Norris}, {Osborne}, {Paczynski}, {Palmer}, {Park}, {Parsons}, {Paul}, {Rees}, {Reynolds}, {Rhoads}, {Sasseen}, {Schaefer}, {Short}, {Smale}, {Smith}, {Stella}, {Tagliaferri}, {Takahashi}, {Tashiro}, {Townsley}, {Tueller}, {Turner}, {Vietri}, {Voges}, {Ward}, {Willingale}, {Zerbi}, \& {Zhang}}]{Gehrels_2004}
{Gehrels}, N., {Chincarini}, G., {Giommi}, P., {et~al.} 2004, \bibinfo{title}{{The Swift Gamma-Ray Burst Mission},} \apj, 611, 1005, \dodoi{10.1086/422091}

\bibitem[{W.~W. {Golay} {et~al.}(2023){Golay}, {Cendes}, {Alexander}, {Berger}, {Eftekhari}, {Pasham}, {Christy}, {Miller-Jones}, \& {Laskar}}]{Golay_2023}
{Golay}, W.~W., {Cendes}, Y., {Alexander}, K.~D., {et~al.} 2023, \bibinfo{title}{{Radio observations of Tidal Disruption Event AT2023mfm},} Transient Name Server AstroNote, 266, 1

\bibitem[{A.~J. {Goodwin} {et~al.}(2025){Goodwin}, {Burn}, {Anderson}, {Miller-Jones}, {Grotova}, {Baldini}, {Liu}, {Malyali}, {Rau}, \& {Salvato}}]{Goodwin_2025}
{Goodwin}, A.~J., {Burn}, M., {Anderson}, G.~E., {et~al.} 2025, \bibinfo{title}{{A Systematic Analysis of the Radio Properties of 22 X-Ray-selected Tidal Disruption Event Candidates with the Australia Telescope Compact Array},} \apjs, 278, 36, \dodoi{10.3847/1538-4365/adbe80}

\bibitem[{A.~J. {Goodwin} {et~al.}(2026){Goodwin}, {Miller-Jones}, {Sfaradi}, {Mummery}, {Margutti}, {Laskar}, {Alexander}, {Yao}, {Balasubramanian}, {Anupama}, {Berger}, {Bhalerao}, {Cendes}, {Chornock}, {Christy}, {Eappachen}, {Eftekhari}, {P{\'e}rez-Torres}, {Ramirez-Ruiz}, {Sahu}, \& {van Velzen}}]{Goodwin_2026}
{Goodwin}, A.~J., {Miller-Jones}, J. C.~A., {Sfaradi}, I., {et~al.} 2026, \bibinfo{title}{{One year of broadband radio monitoring of the enigmatic transient GRB 250702B reveals the evolution of the relativistic jet},} arXiv e-prints, arXiv:2608.03205, \dodoi{10.48550/arXiv.2608.03205}

\bibitem[{M.~J. {Graham} {et~al.}(2019){Graham}, {Kulkarni}, {Bellm}, {Adams}, {Barbarino}, {Blagorodnova}, {Bodewits}, {Bolin}, {Brady}, {Cenko}, {Chang}, {Coughlin}, {De}, {Eadie}, {Farnham}, {Feindt}, {Franckowiak}, {Fremling}, {Gezari}, {Ghosh}, {Goldstein}, {Golkhou}, {Goobar}, {Ho}, {Huppenkothen}, {Ivezi{\'c}}, {Jones}, {Juric}, {Kaplan}, {Kasliwal}, {Kelley}, {Kupfer}, {Lee}, {Lin}, {Lunnan}, {Mahabal}, {Miller}, {Ngeow}, {Nugent}, {Ofek}, {Prince}, {Rauch}, {van Roestel}, {Schulze}, {Singer}, {Sollerman}, {Taddia}, {Yan}, {Ye}, {Yu}, {Barlow}, {Bauer}, {Beck}, {Belicki}, {Biswas}, {Brinnel}, {Brooke}, {Bue}, {Bulla}, {Burruss}, {Connolly}, {Cromer}, {Cunningham}, {Dekany}, {Delacroix}, {Desai}, {Duev}, {Feeney}, {Flynn}, {Frederick}, {Gal-Yam}, {Giomi}, {Groom}, {Hacopians}, {Hale}, {Helou}, {Henning}, {Hover}, {Hillenbrand}, {Howell}, {Hung}, {Imel}, {Ip}, {Jackson}, {Kaspi}, {Kaye}, {Kowalski}, {Kramer}, {Kuhn}, {Landry}, {Laher}, {Mao}, {Masci}, {Monkewitz}, {Murphy}, {Nordin}, {Patterson},
  {Penprase}, {Porter}, {Rebbapragada}, {Reiley}, {Riddle}, {Rigault}, {Rodriguez}, {Rusholme}, {van Santen}, {Shupe}, {Smith}, {Soumagnac}, {Stein}, {Surace}, {Szkody}, {Terek}, {Van Sistine}, {van Velzen}, {Vestrand}, {Walters}, {Ward}, {Zhang}, \& {Zolkower}}]{Graham_2019}
{Graham}, M.~J., {Kulkarni}, S.~R., {Bellm}, E.~C., {et~al.} 2019, \bibinfo{title}{{The Zwicky Transient Facility: Science Objectives},} \pasp, 131, 078001, \dodoi{10.1088/1538-3873/ab006c}

\bibitem[{J.~E. {Greene} {et~al.}(2020){Greene}, {Strader}, \& {Ho}}]{Greene_2020}
{Greene}, J.~E., {Strader}, J., \& {Ho}, L.~C. 2020, \bibinfo{title}{{Intermediate-Mass Black Holes},} \araa, 58, 257, \dodoi{10.1146/annurev-astro-032620-021835}

\bibitem[{I. {Grotova} {et~al.}(2025){Grotova}, {Rau}, {Baldini}, {Goodwin}, {Liu}, {Merloni}, {Salvato}, {Anderson}, {Arcodia}, {Buchner}, {Krumpe}, {Malyali}, {Masterson}, {Miller-Jones}, {Nandra}, \& {Shirley}}]{Grotova_2025}
{Grotova}, I., {Rau}, A., {Baldini}, P., {et~al.} 2025, \bibinfo{title}{{The population of tidal disruption events discovered with eROSITA},} \aap, 697, A159, \dodoi{10.1051/0004-6361/202553669}

\bibitem[{J. {Guillochon} {et~al.}(2018){Guillochon}, {Nicholl}, {Villar}, {Mockler}, {Narayan}, {Mandel}, {Berger}, \& {Williams}}]{Guillochon_2018}
{Guillochon}, J., {Nicholl}, M., {Villar}, V.~A., {et~al.} 2018, \bibinfo{title}{{MOSFiT: Modular Open Source Fitter for Transients},} \apjs, 236, 6, \dodoi{10.3847/1538-4365/aab761}

\bibitem[{M. {Guolo}(2026){Guolo}}]{Guolo_2026}
{Guolo}, M. 2026, \bibinfo{title}{{Demographics of Wandering Black Holes Powering Off-Nuclear Tidal Disruption Events},} arXiv e-prints, arXiv:2602.12970, \dodoi{10.48550/arXiv.2602.12970}

\bibitem[{M. {Guolo} {et~al.}(2024){Guolo}, {Gezari}, {Yao}, {van Velzen}, {Hammerstein}, {Cenko}, \& {Tokayer}}]{Guolo2024}
{Guolo}, M., {Gezari}, S., {Yao}, Y., {et~al.} 2024, \bibinfo{title}{{A Systematic Analysis of the X-Ray Emission in Optically Selected Tidal Disruption Events: Observational Evidence for the Unification of the Optically and X-Ray-selected Populations},} \apj, 966, 160, \dodoi{10.3847/1538-4357/ad2f9f}

\bibitem[{M. {Guolo} {et~al.}(2025){Guolo}, {Mummery}, {van Velzen}, {Gezari}, {Nicholl}, {Yao}, {Karmen}, {Ajay}, {Wevers}, {LeBaron}, \& {Chornock}}]{Guolo_2025}
{Guolo}, M., {Mummery}, A., {van Velzen}, S., {et~al.} 2025, \bibinfo{title}{{Compact Accretion Disks in the Aftermath of Tidal Disruption Events: Parameter Inference from Joint X-ray Spectra and UV/Optical Photometry Fitting},} arXiv e-prints, arXiv:2510.26774, \dodoi{10.48550/arXiv.2510.26774}

\bibitem[{E. {Hammerstein} {et~al.}(2023){Hammerstein}, {van Velzen}, {Gezari}, {Cenko}, {Yao}, {Ward}, {Frederick}, {Villanueva}, {Somalwar}, {Graham}, {Kulkarni}, {Stern}, {Andreoni}, {Bellm}, {Dekany}, {Dhawan}, {Drake}, {Fremling}, {Gatkine}, {Groom}, {Ho}, {Kasliwal}, {Karambelkar}, {Kool}, {Masci}, {Medford}, {Perley}, {Purdum}, {van Roestel}, {Sharma}, {Sollerman}, {Taggart}, \& {Yan}}]{Hammerstein_2023}
{Hammerstein}, E., {van Velzen}, S., {Gezari}, S., {et~al.} 2023, \bibinfo{title}{{The Final Season Reimagined: 30 Tidal Disruption Events from the ZTF-I Survey},} \apj, 942, 9, \dodoi{10.3847/1538-4357/aca283}

\bibitem[{J.~G. {Hills}(1975){Hills}}]{Hills_1975}
{Hills}, J.~G. 1975, \bibinfo{title}{{Possible power source of Seyfert galaxies and QSOs},} \nat, 254, 295, \dodoi{10.1038/254295a0}

\bibitem[{L. {Hoffman} \& A. {Loeb}(2007){Hoffman} \& {Loeb}}]{Hoffman_2007}
{Hoffman}, L., \& {Loeb}, A. 2007, \bibinfo{title}{{Dynamics of triple black hole systems in hierarchically merging massive galaxies},} \mnras, 377, 957, \dodoi{10.1111/j.1365-2966.2007.11694.x}

\bibitem[{J.~D. {Hogg} {et~al.}(2021){Hogg}, {Blecha}, {Reynolds}, {Smith}, \& {Winter}}]{Hogg_2021}
{Hogg}, J.~D., {Blecha}, L., {Reynolds}, C.~S., {Smith}, K.~L., \& {Winter}, L.~M. 2021, \bibinfo{title}{{2MASX J00423991 + 3017515: an offset active galactic nucleus in an interacting system},} \mnras, 503, 1688, \dodoi{10.1093/mnras/stab576}

\bibitem[{C.-C. {Jin} {et~al.}(2025){Jin}, {Li}, {Jiang}, {Dai}, {Cheng}, {Zhu}, {Yang}, {Rau}, {Baldini}, {Wang}, {Zhou}, {Yuan}, {Zhang}, {Shu}, {Shen}, {Wang}, {Wen}, {Wu}, {Wang}, {Thomsen}, {Zhang}, {Zhang}, {Coleiro}, {Eyles-Ferris}, {Fang}, {Ho}, {Hu}, {Jin}, {Li}, {Liu}, {Liu}, {Liu}, {Liu}, {Lu}, {Merloni}, {Qiao}, {Saxton}, {Soria}, {Wang}, {Xue}, {Yang}, {Zhang}, {Zhang}, {Cai}, {Chen}, {Chen}, {Chen}, {Chen}, {Chen}, {Chen}, {Chen}, {Cordier}, {Cui}, {Cui}, {Dai}, {Ding}, {Fan}, {Fan}, {Feng}, {Garcia}, {Guan}, {Han}, {Hou}, {Hu}, {Huang}, {Huo}, {Jia}, {Jia}, {Jiang}, {Jin}, {Kong}, {Kuulkers}, {Lei}, {Li}, {Li}, {Li}, {Li}, {Li}, {Li}, {Lian}, {Ling}, {Liu}, {Liu}, {Liu}, {Liu}, {Liu}, {Lu}, {Luo}, {Ma}, {Mao}, {Mu}, {Nandra}, {O'Brien}, {Pan}, {Pan}, {Qin}, {Rea}, {Sanders}, {Song}, {Sun}, {Sun}, {Sun}, {Tan}, {Tang}, {Tao}, {Wang}, {Wang}, {Wang}, {Wang}, {Wang}, {Wang}, {Wang}, {Wu}, {Wu}, {Xu}, {Xu}, {Xu}, {Xu}, {Xu}, {Xue}, {Xue}, {Xue}, {Yan}, {Yang}, {Yang}, {Zhang}, {Zhang}, {Zhang},
  {Zhang}, {Zhang}, {Zhang}, {Zhang}, {Zhao}, {Zhao}, {Zhao}, {Zhao}, {Zheng}, {Zhu}, {Zhu}, {Zhu}, \& {Zou}}]{Jin_2025}
{Jin}, C.-C., {Li}, D.-Y., {Jiang}, N., {et~al.} 2025, \bibinfo{title}{{An Intermediate-mass Black Hole Lurking in A Galactic Halo Caught Alive during Outburst},} arXiv e-prints, arXiv:2501.09580, \dodoi{10.48550/arXiv.2501.09580}

\bibitem[{B. Johnson {et~al.}(2026)Johnson, Foreman-Mackey, Sick, Leja, Walmsley, Tollerud, Leung, \& Park}]{benjamin_johnson_2026_21778582}
Johnson, B., Foreman-Mackey, D., Sick, J., {et~al.} 2026, dfm/python-fsps: v0.5.0, v0.5.0 Zenodo, \dodoi{10.5281/zenodo.21778582}

\bibitem[{B.~D. {Johnson} {et~al.}(2021){Johnson}, {Leja}, {Conroy}, \& {Speagle}}]{Johnson_2021}
{Johnson}, B.~D., {Leja}, J., {Conroy}, C., \& {Speagle}, J.~S. 2021, \bibinfo{title}{{Stellar Population Inference with Prospector},} \apjs, 254, 22, \dodoi{10.3847/1538-4365/abef67}

\bibitem[{P.~G. {Jonker} {et~al.}(2020){Jonker}, {Stone}, {Generozov}, {van Velzen}, \& {Metzger}}]{Jonker_2020}
{Jonker}, P.~G., {Stone}, N.~C., {Generozov}, A., {van Velzen}, S., \& {Metzger}, B. 2020, \bibinfo{title}{{Implications from Late-time X-Ray Detections of Optically Selected Tidal Disruption Events: State Changes, Unification, and Detection Rates},} \apj, 889, 166, \dodoi{10.3847/1538-4357/ab659c}

\bibitem[{C.~R. {Kaiser}(2006){Kaiser}}]{Kaiser_2006}
{Kaiser}, C.~R. 2006, \bibinfo{title}{{The flat synchrotron spectra of partially self-absorbed jets revisited},} \mnras, 367, 1083, \dodoi{10.1111/j.1365-2966.2006.10030.x}

\bibitem[{L.~J. {Kewley} {et~al.}(2006){Kewley}, {Groves}, {Kauffmann}, \& {Heckman}}]{Kewley_2006}
{Kewley}, L.~J., {Groves}, B., {Kauffmann}, G., \& {Heckman}, T. 2006, \bibinfo{title}{{The host galaxies and classification of active galactic nuclei},} \mnras, 372, 961, \dodoi{10.1111/j.1365-2966.2006.10859.x}

\bibitem[{Y.-L. {Kim} {et~al.}(2022){Kim}, {Rigault}, {Neill}, {Briday}, {Copin}, {Lezmy}, {Nicolas}, {Riddle}, {Sharma}, {Smith}, {Sollerman}, \& {Walters}}]{Kim_2022}
{Kim}, Y.-L., {Rigault}, M., {Neill}, J.~D., {et~al.} 2022, \bibinfo{title}{{New Modules for the SEDMachine to Remove Contaminations from Cosmic Rays and Non-target Light: BYECR and CONTSEP},} \pasp, 134, 024505, \dodoi{10.1088/1538-3873/ac50a0}

\bibitem[{A.~L. {King} {et~al.}(2016){King}, {Miller}, {Bietenholz}, {G{\"u}ltekin}, {Reynolds}, {Mioduszewski}, {Rupen}, \& {Bartel}}]{King_2016}
{King}, A.~L., {Miller}, J.~M., {Bietenholz}, M., {et~al.} 2016, \bibinfo{title}{{Discrete knot ejection from the jet in a nearby low-luminosity active galactic nucleus, M81$^{{\ensuremath{*}}}$},} Nature Physics, 12, 772, \dodoi{10.1038/nphys3724}

\bibitem[{C.~S. {Kochanek} {et~al.}(2017){Kochanek}, {Shappee}, {Stanek}, {Holoien}, {Thompson}, {Prieto}, {Dong}, {Shields}, {Will}, {Britt}, {Perzanowski}, \& {Pojma{\'n}ski}}]{asassn_paper}
{Kochanek}, C.~S., {Shappee}, B.~J., {Stanek}, K.~Z., {et~al.} 2017, \bibinfo{title}{{The All-Sky Automated Survey for Supernovae (ASAS-SN) Light Curve Server v1.0},} \pasp, 129, 104502, \dodoi{10.1088/1538-3873/aa80d9}

\bibitem[{S. {Komossa} \& D. {Merritt}(2008){Komossa} \& {Merritt}}]{Komossa_2008}
{Komossa}, S., \& {Merritt}, D. 2008, \bibinfo{title}{{Tidal Disruption Flares from Recoiling Supermassive Black Holes},} \apjl, 683, L21, \dodoi{10.1086/591420}

\bibitem[{J. {Kormendy} \& L.~C. {Ho}(2013){Kormendy} \& {Ho}}]{Kormendy_2013}
{Kormendy}, J., \& {Ho}, L.~C. 2013, \bibinfo{title}{{Coevolution (Or Not) of Supermassive Black Holes and Host Galaxies},} \araa, 51, 511, \dodoi{10.1146/annurev-astro-082708-101811}

\bibitem[{A.~J. {Levan} {et~al.}(2025){Levan}, {Martin-Carrillo}, {Laskar}, {Eyles-Ferris}, {Sneppen}, {Ravasio}, {Rastinejad}, {Bright}, {Carotenuto}, {Chrimes}, {Corcoran}, {Gompertz}, {Jonker}, {Lamb}, {Malesani}, {Saccardi}, {S{\'a}nchez-Sierras}, {Schneider}, {Schulze}, {Tanvir}, {Vergani}, {Watson}, {An}, {Bauer}, {Campana}, {Cotter}, {van Dalen}, {D'Elia}, {De Pasquale}, {de Ugarte Postigo}, {Dimple}, {Hartmann}, {Hjorth}, {Izzo}, {Jakobsson}, {Kumar}, {Melandri}, {O'Brien}, {Piranomonte}, {Pugliese}, {Quirola-V{\'a}squez}, {Starling}, {Tagliaferri}, {Xu}, \& {Wortley}}]{Levan_2025}
{Levan}, A.~J., {Martin-Carrillo}, A., {Laskar}, T., {et~al.} 2025, \bibinfo{title}{{The Day-long, Repeating GRB 250702B: A Unique Extragalactic Transient},} \apjl, 990, L28, \dodoi{10.3847/2041-8213/adf8e1}

\bibitem[{W. {Li} {et~al.}(2026){Li}, {Christy}, {Alexander}, {Sfaradi}, {Huang}, {Jiang}, {Laskar}, {Mummery}, {Franz}, {Goodwin}, {Golay}, {Margutti}, {Chornock}, {Zhu}, {van Velzen}, {Cendes}, {Lu}, \& {Lynch}}]{Wenkai_2026}
{Li}, W., {Christy}, C.~T., {Alexander}, K.~D., {et~al.} 2026, \bibinfo{title}{{VLA Observations Confirm AT 2023mfm as an Off-nuclear Tidal Disruption Event},} arXiv e-prints, arXiv:2606.06595, \dodoi{10.48550/arXiv.2606.06595}

\bibitem[{D. {Lin} {et~al.}(2018){Lin}, {Strader}, {Carrasco}, {Page}, {Romanowsky}, {Homan}, {Irwin}, {Remillard}, {Godet}, {Webb}, {Baumgardt}, {Wijnands}, {Barret}, {Duc}, {Brodie}, \& {Gwyn}}]{Lin_2018}
{Lin}, D., {Strader}, J., {Carrasco}, E.~R., {et~al.} 2018, \bibinfo{title}{{A luminous X-ray outburst from an intermediate-mass black hole in an off-centre star cluster},} Nature Astronomy, 2, 656, \dodoi{10.1038/s41550-018-0493-1}

\bibitem[{D. {Lin} {et~al.}(2020){Lin}, {Strader}, {Romanowsky}, {Irwin}, {Godet}, {Barret}, {Webb}, {Homan}, \& {Remillard}}]{Lin_2020}
{Lin}, D., {Strader}, J., {Romanowsky}, A.~J., {et~al.} 2020, \bibinfo{title}{{Multiwavelength Follow-up of the Hyperluminous Intermediate-mass Black Hole Candidate 3XMM J215022.4-055108},} \apjl, 892, L25, \dodoi{10.3847/2041-8213/ab745b}

\bibitem[{J.~M. {Lotz} {et~al.}(2008){Lotz}, {Jonsson}, {Cox}, \& {Primack}}]{Lotz_2008}
{Lotz}, J.~M., {Jonsson}, P., {Cox}, T.~J., \& {Primack}, J.~R. 2008, \bibinfo{title}{{Galaxy merger morphologies and time-scales from simulations of equal-mass gas-rich disc mergers},} \mnras, 391, 1137, \dodoi{10.1111/j.1365-2966.2008.14004.x}

\bibitem[{B. {Magnelli} {et~al.}(2015){Magnelli}, {Ivison}, {Lutz}, {Valtchanov}, {Farrah}, {Berta}, {Bertoldi}, {Bock}, {Cooray}, {Ibar}, {Karim}, {Le Floc'h}, {Nordon}, {Oliver}, {Page}, {Popesso}, {Pozzi}, {Rigopoulou}, {Riguccini}, {Rodighiero}, {Rosario}, {Roseboom}, {Wang}, \& {Wuyts}}]{Magnelli_2015}
{Magnelli}, B., {Ivison}, R.~J., {Lutz}, D., {et~al.} 2015, \bibinfo{title}{{The far-infrared/radio correlation and radio spectral index of galaxies in the SFR-M$_{{\ensuremath{*}}}$ plane up to z\raisebox{-0.5ex}\textasciitilde2},} \aap, 573, A45, \dodoi{10.1051/0004-6361/201424937}

\bibitem[{F.~J. {Masci} {et~al.}(2019){Masci}, {Laher}, {Rusholme}, {Shupe}, {Groom}, {Surace}, {Jackson}, {Monkewitz}, {Beck}, {Flynn}, {Terek}, {Landry}, {Hacopians}, {Desai}, {Howell}, {Brooke}, {Imel}, {Wachter}, {Ye}, {Lin}, {Cenko}, {Cunningham}, {Rebbapragada}, {Bue}, {Miller}, {Mahabal}, {Bellm}, {Patterson}, {Juri{\'c}}, {Golkhou}, {Ofek}, {Walters}, {Graham}, {Kasliwal}, {Dekany}, {Kupfer}, {Burdge}, {Cannella}, {Barlow}, {Van Sistine}, {Giomi}, {Fremling}, {Blagorodnova}, {Levitan}, {Riddle}, {Smith}, {Helou}, {Prince}, \& {Kulkarni}}]{Masci_2019}
{Masci}, F.~J., {Laher}, R.~R., {Rusholme}, B., {et~al.} 2019, \bibinfo{title}{{The Zwicky Transient Facility: Data Processing, Products, and Archive},} \pasp, 131, 018003, \dodoi{10.1088/1538-3873/aae8ac}

\bibitem[{F.~J. {Masci} {et~al.}(2023){Masci}, {Laher}, {Rusholme}, {Shupe}, {Paladini}, {Groom}, {Wold}, {Miller}, \& {Drake}}]{Masci_2023}
{Masci}, F.~J., {Laher}, R.~R., {Rusholme}, B., {et~al.} 2023, \bibinfo{title}{{A New Forced Photometry Service for the Zwicky Transient Facility},} arXiv e-prints, arXiv:2305.16279, \dodoi{10.48550/arXiv.2305.16279}

\bibitem[{T. {Matsumoto} \& T. {Piran}(2023){Matsumoto} \& {Piran}}]{Matsumoto_2023}
{Matsumoto}, T., \& {Piran}, T. 2023, \bibinfo{title}{{Generalized equipartition method from an arbitrary viewing angle},} \mnras, 522, 4565, \dodoi{10.1093/mnras/stad1269}

\bibitem[{J.~S. Miller(1994)Miller}]{Miller_1994}
Miller, J.~S. 1994, The {{Kast Double Spectograph}} (University of California Observatories/Lick Observatory)

\bibitem[{B. {Mockler} {et~al.}(2019){Mockler}, {Guillochon}, \& {Ramirez-Ruiz}}]{Mockler_2019}
{Mockler}, B., {Guillochon}, J., \& {Ramirez-Ruiz}, E. 2019, \bibinfo{title}{{Weighing Black Holes Using Tidal Disruption Events},} \apj, 872, 151, \dodoi{10.3847/1538-4357/ab010f}

\bibitem[{A. {Mummery} \& S.~A. {Balbus}(2020){Mummery} \& {Balbus}}]{Mummery_2020}
{Mummery}, A., \& {Balbus}, S.~A. 2020, \bibinfo{title}{{The spectral evolution of disc dominated tidal disruption events},} \mnras, 492, 5655, \dodoi{10.1093/mnras/staa192}

\bibitem[{A. {Mummery} {et~al.}(2024){Mummery}, {van Velzen}, {Nathan}, {Ingram}, {Hammerstein}, {Fraser-Taliente}, \& {Balbus}}]{Mummery_2024}
{Mummery}, A., {van Velzen}, S., {Nathan}, E., {et~al.} 2024, \bibinfo{title}{{Fundamental scaling relationships revealed in the optical light curves of tidal disruption events},} \mnras, 527, 2452, \dodoi{10.1093/mnras/stad3001}

\bibitem[{N. {Neumayer} {et~al.}(2020){Neumayer}, {Seth}, \& {B{\"o}ker}}]{Neumayer_2020}
{Neumayer}, N., {Seth}, A., \& {B{\"o}ker}, T. 2020, \bibinfo{title}{{Nuclear star clusters},} \aapr, 28, 4, \dodoi{10.1007/s00159-020-00125-0}

\bibitem[{J.~B. {Oke} {et~al.}(1995){Oke}, {Cohen}, {Carr}, {Cromer}, {Dingizian}, {Harris}, {Labrecque}, {Lucinio}, {Schaal}, {Epps}, \& {Miller}}]{Oke_1995}
{Oke}, J.~B., {Cohen}, J.~G., {Carr}, M., {et~al.} 1995, \bibinfo{title}{{The Keck Low-Resolution Imaging Spectrometer},} \pasp, 107, 375, \dodoi{10.1086/133562}

\bibitem[{K.~C. {Patra} {et~al.}(2025){Patra}, {Foley}, {Earl}, {Davis}, {Ramirez-Ruiz}, {Villar}, {Gomez}, {French}, {Taggart}, {Arunachalam}, {Macias}, {Kaur}, \& {Tinyanont}}]{Patra_2025}
{Patra}, K.~C., {Foley}, R.~J., {Earl}, N., {et~al.} 2025, \bibinfo{title}{{JWST and Keck Observations of the Off-Nuclear TDE AT 2024tvd: A Massive Nuclear Star Cluster and Minor-Merger Origin for its Black Hole},} arXiv e-prints, arXiv:2510.12572, \dodoi{10.48550/arXiv.2510.12572}

\bibitem[{K.~C. {Patra} {et~al.}(2026){Patra}, {Liepold}, {Earl}, {Foley}, {Ma}, {Gomez}, {Davis}, {Ramirez-Ruiz}, {French}, {Walsh}, {Kaur}, {Taggart}, {Candanoza}, {Villar}, {Arunachalam}, {Macias}, \& {Tinyanont}}]{Patra_2026}
{Patra}, K.~C., {Liepold}, E.~R., {Earl}, N., {et~al.} 2026, \bibinfo{title}{{JWST and Keck observations of the off-nuclear tidal disruption event TDE 2025abcr: An evolving reprocessing layer},} arXiv e-prints, arXiv:2604.16093.
\newblock \doarXiv{2604.16093}

\bibitem[{F. Pedregosa {et~al.}(2011)Pedregosa, Varoquaux, Gramfort, Michel, Thirion, Grisel, Blondel, Prettenhofer, Weiss, Dubourg, Vanderplas, Passos, Cournapeau, Brucher, Perrot, \& Duchesnay}]{scikit-learn}
Pedregosa, F., Varoquaux, G., Gramfort, A., {et~al.} 2011, \bibinfo{title}{Scikit-learn: Machine Learning in {P}ython,} Journal of Machine Learning Research, 12, 2825

\bibitem[{M.~J. {Rees}(1988){Rees}}]{Rees_1988}
{Rees}, M.~J. 1988, \bibinfo{title}{{Tidal disruption of stars by black holes of {}10$^{6}$-{}10$^{8}$ solar masses in nearby galaxies},} \nat, 333, 523, \dodoi{10.1038/333523a0}

\bibitem[{A. {Ricarte} {et~al.}(2021{\natexlab{a}}){Ricarte}, {Tremmel}, {Natarajan}, \& {Quinn}}]{Ricarte_2021_EM}
{Ricarte}, A., {Tremmel}, M., {Natarajan}, P., \& {Quinn}, T. 2021{\natexlab{a}}, \bibinfo{title}{{Unveiling the Population of Wandering Black Holes via Electromagnetic Signatures},} \apjl, 916, L18, \dodoi{10.3847/2041-8213/ac1170}

\bibitem[{A. {Ricarte} {et~al.}(2021{\natexlab{b}}){Ricarte}, {Tremmel}, {Natarajan}, {Zimmer}, \& {Quinn}}]{Ricarte_2021_demographic}
{Ricarte}, A., {Tremmel}, M., {Natarajan}, P., {Zimmer}, C., \& {Quinn}, T. 2021{\natexlab{b}}, \bibinfo{title}{{Origins and demographics of wandering black holes},} \mnras, 503, 6098, \dodoi{10.1093/mnras/stab866}

\bibitem[{M. {Rigault} {et~al.}(2019){Rigault}, {Neill}, {Blagorodnova}, {Dugas}, {Feeney}, {Walters}, {Brinnel}, {Copin}, {Fremling}, {Nordin}, \& {Sollerman}}]{Rigault_2019}
{Rigault}, M., {Neill}, J.~D., {Blagorodnova}, N., {et~al.} 2019, \bibinfo{title}{{Fully automated integral field spectrograph pipeline for the SEDMachine: pysedm},} \aap, 627, A115, \dodoi{10.1051/0004-6361/201935344}

\bibitem[{P.~W.~A. {Roming} {et~al.}(2005){Roming}, {Kennedy}, {Mason}, {Nousek}, {Ahr}, {Bingham}, {Broos}, {Carter}, {Hancock}, {Huckle}, {Hunsberger}, {Kawakami}, {Killough}, {Koch}, {McLelland}, {Smith}, {Smith}, {Soto}, {Boyd}, {Breeveld}, {Holland}, {Ivanushkina}, {Pryzby}, {Still}, \& {Stock}}]{Roming_2005}
{Roming}, P. W.~A., {Kennedy}, T.~E., {Mason}, K.~O., {et~al.} 2005, \bibinfo{title}{{The Swift Ultra-Violet/Optical Telescope},} \ssr, 120, 95, \dodoi{10.1007/s11214-005-5095-4}

\bibitem[{T. {Ryu} {et~al.}(2018){Ryu}, {Perna}, {Haiman}, {Ostriker}, \& {Stone}}]{Ryu_2018}
{Ryu}, T., {Perna}, R., {Haiman}, Z., {Ostriker}, J.~P., \& {Stone}, N.~C. 2018, \bibinfo{title}{{Interactions between multiple supermassive black holes in galactic nuclei: a solution to the final parsec problem},} \mnras, 473, 3410, \dodoi{10.1093/mnras/stx2524}

\bibitem[{R. {Sari} {et~al.}(1998){Sari}, {Piran}, \& {Narayan}}]{Sari_1998}
{Sari}, R., {Piran}, T., \& {Narayan}, R. 1998, \bibinfo{title}{{Spectra and Light Curves of Gamma-Ray Burst Afterglows},} \apjl, 497, L17, \dodoi{10.1086/311269}

\bibitem[{E.~F. {Schlafly} \& D.~P. {Finkbeiner}(2011){Schlafly} \& {Finkbeiner}}]{Schlafly_2011}
{Schlafly}, E.~F., \& {Finkbeiner}, D.~P. 2011, \bibinfo{title}{{Measuring Reddening with Sloan Digital Sky Survey Stellar Spectra and Recalibrating SFD},} \apj, 737, 103, \dodoi{10.1088/0004-637X/737/2/103}

\bibitem[{H. {Sears} {et~al.}(2026){Sears}, {Somalwar}, {Chornock}, {Laskar}, {Levan}, {Margutti}, {O'Connor}, {Nayana A.}, {Ahumada}, {Alexander}, {Andreoni}, {Anumarlapudi}, {Carney}, {Freeburn}, {Galbany}, {Gompertz}, {Graur}, {Hall}, {Hall}, {Hammerstein}, {Jha}, {Kasliwal}, {Pasham}, {Sfaradi}, \& {Yao}}]{Sears_2026}
{Sears}, H., {Somalwar}, J.~J., {Chornock}, R., {et~al.} 2026, \bibinfo{title}{{Late-time JWST/NIRCam Observations of the Extremely Long-duration GRB 250702B/EP 250702a and Its Host Galaxy},} arXiv e-prints, arXiv:2606.18353, \dodoi{10.48550/arXiv.2606.18353}

\bibitem[{I. {Sfaradi} {et~al.}(2022){Sfaradi}, {Horesh}, {Fender}, {Green}, {Williams}, {Bright}, \& {Schulze}}]{Sfaradi_2022}
{Sfaradi}, I., {Horesh}, A., {Fender}, R., {et~al.} 2022, \bibinfo{title}{{A Late-time Radio Flare Following a Possible Transition in Accretion State in the Tidal Disruption Event AT 2019azh},} \apj, 933, 176, \dodoi{10.3847/1538-4357/ac74bc}

\bibitem[{I. {Sfaradi} {et~al.}(2024){Sfaradi}, {Beniamini}, {Horesh}, {Piran}, {Bright}, {Rhodes}, {Williams}, {Fender}, {Leung}, {Murphy}, \& {Green}}]{Sfaradi_2024}
{Sfaradi}, I., {Beniamini}, P., {Horesh}, A., {et~al.} 2024, \bibinfo{title}{{An off-axis relativistic jet seen in the long lasting delayed radio flare of the TDE AT 2018hyz},} \mnras, 527, 7672, \dodoi{10.1093/mnras/stad3717}

\bibitem[{I. {Sfaradi} {et~al.}(2025){Sfaradi}, {Margutti}, {Chornock}, {Alexander}, {Metzger}, {Beniamini}, {Duran}, {Yao}, {Horesh}, {Farah}, {Berger}, {A.~J.}, {Cendes}, {Eftekhari}, {Fender}, {Franz}, {Green}, {Hammerstein}, {Lu}, {Wiston}, {Bernstein}, {Bright}, {Christy}, {Cruz}, {DeBoer}, {Golay}, {Goodwin}, {Gurwell}, {Keating}, {Laskar}, {Miller-Jones}, {Pollak}, {Rao}, {Siemion}, {Sheikh}, {Shoval}, \& {van Velzen}}]{Sfaradi_2025}
{Sfaradi}, I., {Margutti}, R., {Chornock}, R., {et~al.} 2025, \bibinfo{title}{{The First Radio-bright Off-nuclear Tidal Disruption Event AT 2024tvd Reveals the Fastest-evolving Double-peaked Radio Emission},} \apjl, 992, L18, \dodoi{10.3847/2041-8213/ae0a26}

\bibitem[{R. {Stein} {et~al.}(2026){Stein}, {Carney}, {Ward}, {Margutti}, {Hall}, {Sfaradi}, {Andreoni}, {Charalampopoulos}, {Chornock}, {Gezari}, {Mo}, {Yao}, {Anumarlapudi}, {Bellm}, {Bloom}, {Busmann}, {Caiazzo}, {Cenko}, {Graham}, {Groom}, {Gruen}, {Hammerstein}, {Kaiser}, {Kasliwal}, {O'Connor}, {Palmese}, {Purdum}, {Rastinejad}, {Riddle}, {Rusholme}, {Sollerman}, {Somalwar}, \& {Veilleux}}]{Stein_2026}
{Stein}, R., {Carney}, J., {Ward}, C., {et~al.} 2026, \bibinfo{title}{{TDE 2025abcr: A Tidal Disruption Event in the Outskirts of a Massive Galaxy},} \apjl, 1006, L57, \dodoi{10.3847/2041-8213/ae77f3}

\bibitem[{N. {Stone} \& A. {Loeb}(2012){Stone} \& {Loeb}}]{Stone_2012}
{Stone}, N., \& {Loeb}, A. 2012, \bibinfo{title}{{Tidal disruption flares of stars from moderately recoiled black holes},} \mnras, 422, 1933, \dodoi{10.1111/j.1365-2966.2012.20577.x}

\bibitem[{C. {Tadhunter}(2016){Tadhunter}}]{Tadhunter_2016}
{Tadhunter}, C. 2016, \bibinfo{title}{{Radio AGN in the local universe: unification, triggering and evolution},} \aapr, 24, 10, \dodoi{10.1007/s00159-016-0094-x}

\bibitem[{M. {Tremmel} {et~al.}(2018{\natexlab{a}}){Tremmel}, {Governato}, {Volonteri}, {Pontzen}, \& {Quinn}}]{Tremmel_2018_wandering}
{Tremmel}, M., {Governato}, F., {Volonteri}, M., {Pontzen}, A., \& {Quinn}, T.~R. 2018{\natexlab{a}}, \bibinfo{title}{{Wandering Supermassive Black Holes in Milky-Way-mass Halos},} \apjl, 857, L22, \dodoi{10.3847/2041-8213/aabc0a}

\bibitem[{M. {Tremmel} {et~al.}(2018{\natexlab{b}}){Tremmel}, {Governato}, {Volonteri}, {Quinn}, \& {Pontzen}}]{Tremmel_2018_pair}
{Tremmel}, M., {Governato}, F., {Volonteri}, M., {Quinn}, T.~R., \& {Pontzen}, A. 2018{\natexlab{b}}, \bibinfo{title}{{Dancing to CHANGA: a self-consistent prediction for close SMBH pair formation time-scales following galaxy mergers},} \mnras, 475, 4967, \dodoi{10.1093/mnras/sty139}

\bibitem[{S.~J. van~der Walt {et~al.}(2019)van~der Walt, Crellin-Quick, \& Bloom}]{van_der_Walt2019}
van~der Walt, S.~J., Crellin-Quick, A., \& Bloom, J.~S. 2019, \bibinfo{title}{SkyPortal: An Astronomical Data Platform,} Journal of Open Source Software, 4, 1247, \dodoi{10.21105/joss.01247}

\bibitem[{S. {van Velzen} {et~al.}(2021){van Velzen}, {Gezari}, {Hammerstein}, {Roth}, {Frederick}, {Ward}, {Hung}, {Cenko}, {Stein}, {Perley}, {Taggart}, {Foley}, {Sollerman}, {Blagorodnova}, {Andreoni}, {Bellm}, {Brinnel}, {De}, {Dekany}, {Feeney}, {Fremling}, {Giomi}, {Golkhou}, {Graham}, {Ho}, {Kasliwal}, {Kilpatrick}, {Kulkarni}, {Kupfer}, {Laher}, {Mahabal}, {Masci}, {Miller}, {Nordin}, {Riddle}, {Rusholme}, {van Santen}, {Sharma}, {Shupe}, \& {Soumagnac}}]{van_velzen_2021}
{van Velzen}, S., {Gezari}, S., {Hammerstein}, E., {et~al.} 2021, \bibinfo{title}{{Seventeen Tidal Disruption Events from the First Half of ZTF Survey Observations: Entering a New Era of Population Studies},} \apj, 908, 4, \dodoi{10.3847/1538-4357/abc258}

\bibitem[{S. {Van Wassenhove} {et~al.}(2012){Van Wassenhove}, {Volonteri}, {Mayer}, {Dotti}, {Bellovary}, \& {Callegari}}]{Van_Wassenhove_2012}
{Van Wassenhove}, S., {Volonteri}, M., {Mayer}, L., {et~al.} 2012, \bibinfo{title}{{Observability of Dual Active Galactic Nuclei in Merging Galaxies},} \apjl, 748, L7, \dodoi{10.1088/2041-8205/748/1/L7}

\bibitem[{A. {Vazdekis} {et~al.}(2016){Vazdekis}, {Koleva}, {Ricciardelli}, {R{\"o}ck}, \& {Falc{\'o}n-Barroso}}]{Vazdekis_2016}
{Vazdekis}, A., {Koleva}, M., {Ricciardelli}, E., {R{\"o}ck}, B., \& {Falc{\'o}n-Barroso}, J. 2016, \bibinfo{title}{{UV-extended E-MILES stellar population models: young components in massive early-type galaxies},} \mnras, 463, 3409, \dodoi{10.1093/mnras/stw2231}

\bibitem[{Y. {Yao} {et~al.}(2023){Yao}, {Ravi}, {Gezari}, {van Velzen}, {Lu}, {Schulze}, {Somalwar}, {Kulkarni}, {Hammerstein}, {Nicholl}, {Graham}, {Perley}, {Cenko}, {Stein}, {Ricarte}, {Chadayammuri}, {Quataert}, {Bellm}, {Bloom}, {Dekany}, {Drake}, {Groom}, {Mahabal}, {Prince}, {Riddle}, {Rusholme}, {Sharma}, {Sollerman}, \& {Yan}}]{Yao_2023}
{Yao}, Y., {Ravi}, V., {Gezari}, S., {et~al.} 2023, \bibinfo{title}{{Tidal Disruption Event Demographics with the Zwicky Transient Facility: Volumetric Rates, Luminosity Function, and Implications for the Local Black Hole Mass Function},} \apjl, 955, L6, \dodoi{10.3847/2041-8213/acf216}

\bibitem[{Y. {Yao} {et~al.}(2025){Yao}, {Chornock}, {Ward}, {Hammerstein}, {Sfaradi}, {Margutti}, {Kelley}, {Lu}, {Liu}, {Wise}, {Sollerman}, {Alexander}, {Bellm}, {Drake}, {Fremling}, {Gilfanov}, {Graham}, {Groom}, {Hinds}, {Kulkarni}, {Miller}, {Miller-Jones}, {Nicholl}, {Perley}, {Purdum}, {Ravi}, {Rich}, {Rehemtulla}, {Riddle}, {Smith}, {Stein}, {Sunyaev}, {van Velzen}, \& {Wold}}]{Yao_2025}
{Yao}, Y., {Chornock}, R., {Ward}, C., {et~al.} 2025, \bibinfo{title}{{A Massive Black Hole 0.8 kpc from the Host Nucleus Revealed by the Offset Tidal Disruption Event AT2024tvd},} \apjl, 985, L48, \dodoi{10.3847/2041-8213/add7de}

\bibitem[{F. {Zou} {et~al.}(2025){Zou}, {Gallo}, {Seth}, {Hodges-Kluck}, {Ohlson}, {Treu}, {Baldassare}, {Brandt}, {Greene}, {Madau}, {Nguyen}, {Plotkin}, {Reines}, {Sesana}, {Woo}, \& {Wu}}]{Zou2025}
{Zou}, F., {Gallo}, E., {Seth}, A.~C., {et~al.} 2025, \bibinfo{title}{{Central Massive Black Holes Are Not Ubiquitous in Local Low-mass Galaxies},} \apj, 992, 176, \dodoi{10.3847/1538-4357/ae06a1}

\end{thebibliography}
\bibliographystyle{aasjournalv7}

\end{document}